\documentclass[a4paper,review]{cas-dc}

\usepackage[authoryear]{natbib}
\usepackage{tabularx}
\usepackage{caption}
\usepackage{amsfonts,amssymb,amsmath,amsgen,amsopn,amsbsy,theorem,graphicx,epsfig, lscape, threeparttable, rotating,tabularx, adjustbox,listings}

\usepackage[ruled,vlined]{algorithm2e}

\usepackage[T1]{fontenc}
\usepackage{rotating}
\usepackage{booktabs}  
\usepackage{siunitx}  

\definecolor{codegreen}{rgb}{0,0.6,0}
\definecolor{codegray}{rgb}{0.5,0.5,0.5}
\definecolor{codepurple}{rgb}{0.58,0,0.82}
\definecolor{backcolour}{rgb}{0.95,0.95,0.92}

\lstdefinestyle{mystyle}{
  backgroundcolor=\color{backcolour},   commentstyle=\color{codegreen},
  keywordstyle=\color{magenta},
  numberstyle=\tiny\color{codegray},
  stringstyle=\color{codepurple},
  basicstyle=\ttfamily\footnotesize,
  breakatwhitespace=false,         
  breaklines=true,                 
  captionpos=b,                    
  keepspaces=true,                 
  numbers=left,                    
  numbersep=5pt,                  
  showspaces=false,                
  showstringspaces=false,
  showtabs=false,                  
  tabsize=2
}
\def\tsc#1{\csdef{#1}{\textsc{\lowercase{#1}}\xspace}}
\tsc{WGM}
\tsc{QE}
\tsc{EP}
\tsc{PMS}
\tsc{BEC}
\tsc{DE}

\begin{document}\sloppy
\let\WriteBookmarks\relax
\def\floatpagepagefraction{1}
\def\textpagefraction{.001}
\shorttitle{Unsupervised Behaviour Analysis of News Consumption in Turkish Media} 
\shortauthors{Makaro\u{g}lu et~al.}
\title [mode = title]{Unsupervised Behaviour Analysis of News Consumption in Turkish Media}                  

\author[1,2]{Didem Makaro\u{g}lu}[orcid=0000-0001-9960-2113]
\cormark[1]
\ead{didem.makaroglu@demirorenteknoloji.com}
\ead{makaroglu17@itu.edu.tr}

\credit{Conceptualization, Methodology, Software, Validation, Investigation, Data curation, Writing - original draft, Writing - review and editing, Visualization}

\address[1]{Demir\"{o}ren Teknoloji A.S., İstanbul, Turkey}
\address[2]{Signal Processing for Computational Intelligence Group, Informatics Institute, Istanbul Technical University, Istanbul, Turkey}

\author[3,4]{Altan Cakir}[orcid=0000-0002-8627-7689]
\ead{altan.cakir@itu.edu.tr}

\credit{Conceptualization of this study, supervising, review and editing}
\address[3]{Physics Engineering, Faculty of Science and Letters, Istanbul Technical University, Istanbul, Turkey}
\address[4]{Istanbul Technical University Artificial Intelligence, Data Science Research and Application Center, Istanbul, Turkey}

\author[2]{Beh\c{c}et U\u{g}ur T\"{o}reyin}[orcid=0000-0003-4406-2783]
\ead{toreyin@itu.edu.tr}
\credit{Conceptualization of this study, Methodology, review and editing}
\cortext[cor1]{Corresponding author}

\begin{abstract}
Clickstream data, which come with a massive volume generated by human activities on websites, have become a prominent feature for identifying readers’ characteristics by newsrooms after the digitization of news outlets. It is essential to have elastic architectures to process the streaming data, particularly for unprecedented traffic, enabling conducting more comprehensive analyses, such as recommending mostly related articles to readers. Although the nature of clickstream data has a similar logic within websites, it has inherent limitations to recognize human behaviours when looking from a broad perspective, which brings the need to limit the problem in niche areas. This study investigates the anonymized readers’ click activities on the organizations’ websites to identify news consumption patterns following referrals from~Twitter, who incidentally reach but propensity is mainly routed news content. The investigation is widened to a broad perspective by linking the log data with news content to enrich the insights rather than sticking into the web journey. Methodologies for ensemble cluster analysis with mixed-type embedding strategies are applied and compared to find similar reader groups and interests independent of time. Various internal validation perspectives are used to determine the optimality of the quality of clusters, where the Calinski Harabasz Index (CHI) is found to give a generalizable result. Our findings demonstrate that clustering a mixed-type dataset approaches the optimal internal validation scores, which we define to discriminate the clusters and algorithms considering applied strategies, when embedded by Uniform Manifold Approximation and Projection (UMAP) and using a consensus function as a key to access the most applicable hyper parameter configurations in the given ensemble rather than using consensus function results directly. Evaluation of the resulting clusters highlights specific clusters repeatedly present in the separated monthly samples by Adjusted Mutual Information scores greater than 0.5, which provide insights to the news organizations and overcome the degradation of the modelling behaviours due to the change in the interest over time.

\end{abstract}


\maketitle
\section{Introduction}
In recent years, news readership behaviours have shifted towards online channels after the digitization of news organizations and the rise of social media ~\citep{consiumingnews_Flaxman_2016}, which has a growing effect on consuming news. The importance of acknowledging the trend of news sources has been aroused by news outlets that need to generate new revenue to adapt and survive in this era. Relatedly, the cruciality of intermediaries has been recognized, and early studies have focused on identifying the effects of these sources on attracting and holding consumers’ attention by guiding users to new content~\citep{Benlian, Tucker, Dellarocas}. Recent studies have shown that reaching news content through social and nonsocial channels shows different patterns~\citep{Moller}, how news reader behaviours are formed with social (Facebook) and nonsocial media channel referrals at an aggregated level~\citep{Bar-Gill}, and how readers spend time on the website and have the propensity to share the news differentiated by the referral channels~\citep{Koster}. As a microservice,~Twitter has become a key player in social media and is used more than on average across 37 countries for reading, sharing, and discussing news in Turkey~\citep{TurkeyReport}. Strongly preferred as disseminating the news,~Twitter also mediates access to pages of news outlets. Considering that each social media site has different page structures, dynamics, and methods of influencing users, it is necessary to examine them separately to analyse the readers directed from these channels to news outlets. To the best of our knowledge, using Twitter as a social referring channel in identifying the behavioural patterns of the readers directed to the news organizations' websites by clicking on news links on Twitter, have yet to be studied. In this paper, we delve into uncovering referred users’ behaviour by clustering methodologies. However, readers tend to navigate away from social media and direct towards news websites with broad and diverse interests and intentions, making it difficult to cluster for long periods. Therefore, we hypothesized and experimentally showed that working with separated datasets in this multidimensional environment is advantageous in recognizing and generalizing user patterns. By providing the distribution of the news category interest breakdowns of individual and general behaviours over time, we found that the behaviours diverge over months; hence, working monthly datasets is beneficial in diminishing the volume and increasing the correlation within the samples.

\section{Research Setting} 
The investigated datasets belong to H\"{u}rriyet, a news organization located in Turkey, founded in 1948 and is active not only on its website since 2004 but also promotes the contents and live broadcast in its social media accounts. In our one-year period of analysis, between April 2019 and March 2020, the website had approximately 150 M unique users and 300 M visits per month. The traffic from the reference sites can be grouped as Social (Facebook,~Twitter, Instagram, LinkedIn, YouTube) and Nonsocial (Organic, referred by Google, other), with monthly visit shares of 8\% and 92\%, respectively, and~Twitter has 25\% of all social referrals. The objective of this study is to analyse referral users’ behaviour; thus, within this scope, our datasets include clickstream data, which are collected from the organization’s website, and news content data. We limit our study with the collected data to uncover unsupervised behaviours of the referred users’ patterns without comparing the findings with the nonsocial referrals, so this study measures neither the contribution of social referral users’ behaviour and news consumption on the website nor the causal effect of being directed to the news outlets~\citep{Mao} . This work presents a comprehensive methodology of ensemble clustering and includes different kinds of embedding as input (i.e., UMAP) after experimenting with different sampling strategies (i.e., random, stratify). Our method evaluates performance metrics with clustering techniques to ensure that behavioural clusters are stable and optimal within this structure. Furthermore, we provide a validation process to measure the similarity of the ensemble clustering with the optimal results. We evaluate the clusters generated by the selected clustering method over twelve months and show the behavioural patterns’ stability and consistency in specific clusters.

\begin{figure*}
    \centering
    \includegraphics[width=17cm, height=10cm]{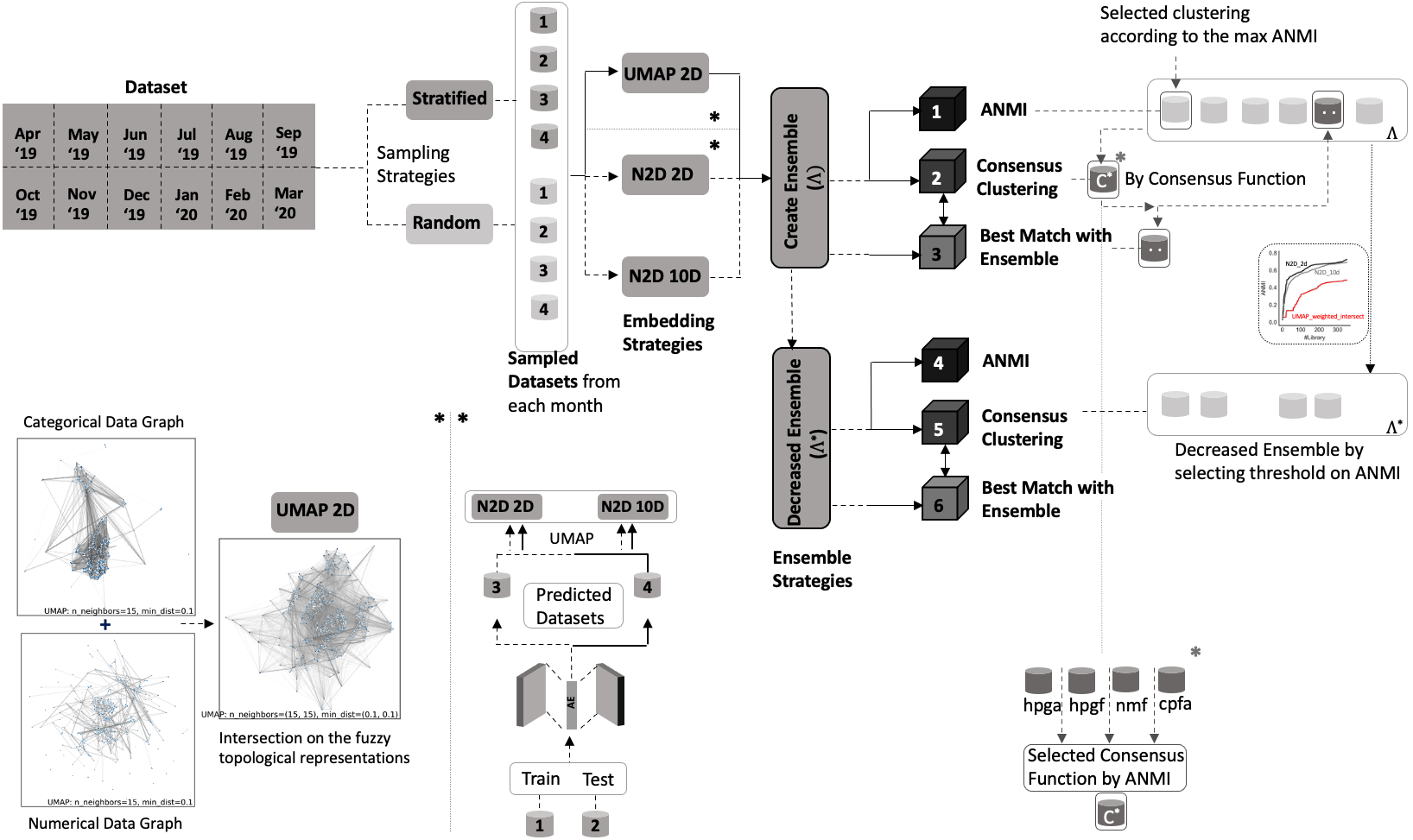}
    
\caption{Methodology architecture. N2d embedding strategies preparation is shown at the bottom-left corner. With the N2d embedding approach, the $1^{st}$ and $2^{nd}$ samples are predicted by training on the $3^{rd}$ and $4^{th}$ samples. UMAP 2D is prepared by intersecting categorical and numerical feature graphs as illustrated at the bottom-left corner for one sampled dataset. The AAMI values are calculated for each consensus function result and then selected with the maximum value, shown at the bottom-right corner.}  
    \label{fig:model_architecture}
\end{figure*}

The rest of the paper is organized as follows and can be followed from Figure~\ref{fig:model_architecture}, which illustrates the framework of our system. Section~\ref{Literature} provides a discussion about related research on clickstream analysis and modelling, clustering with input data-types. Section~\ref{DataAnalysis} presents our dataset, including the collection and processing data platform and the analysis of the distributions. Section~\ref{Methodology} describes the methodologies utilized in this work, including sampling and embedding strategies, ensemble clustering evaluation and comparisons, and discusses successful clustering. Finally, Section~\ref{conclusion} concludes the article and presents our future work.

\section{Related Literature} \label{Literature}
Clickstream data have been used to identify web usage and characteristics in many contexts~\citep{Catledge, Srivastava}. Clustering techniques have been studied to utilize hidden user behaviours~\citep{SuChen, WangKonolige, DanielGrech} and related to this approach, unsupervised techniques have been established to predict future groups~\citep{ZhangKamps}. To identify the dominant behaviour groups, social networks and time-based frameworks are proposed~\citep{lin2014personalized} and enriched by having similarities to the group behaviour and individual propensities~\citep{zheng2013penetrate} . As news readers have been studied in various approaches, recently, the research focus shifted to differentiating news consumption by considering the effects of referral channels~\citep{Moller, Koster} and intermediaries’ power on the algorithmic level~\citep{ThorsonWells, WellsChris}. In our study, we also aim to find unsupervised readership behaviours on the news outlets for the specific referral channel~Twitter by using ensemble clustering methodologies without having neither ground truth nor predefined patterns in the data~\citep{WangZhang, bogaard2019searching}.

As an unsupervised approach, for clustering user behaviours, there are different techniques, including centroid-based (K-means, K-Medoids)~\citep{centroid-based}, hierarchy-based (Agglomerative) and density-based (DBSCAN~\citep{dbscan}, HDBSCAN~\citep{hdbscan} ) partitions. These algorithms have some advantages and disadvantages when considering various factors, such as scalability, time and space complexity, the ability to deal with categorical and numerical data types, noise and outliers. These algorithms need predefined values, including $k$ numbers (K-means, BIRCH, Gaussian Mixture), the epsilon value, the minimum number of points (DBSCAN, HDBSCAN), and linkage metrics including average, single, complete, or ward (Agglomerative). In recent years, artificial neural networks have also been used for unsupervised clustering methods by applying autoencoders. One method is harnessed a deep autoencoder architecture~\citep{N2D} we refer to as N2d, which takes advantage of deep neural networks to represent lower-dimensional data. With this method, after learning autoencoder embedding representations, UMAP~\citep{umap} is applied to better clustering with lower dimensions using local manifold learning. Similar to other clustering algorithms, this algorithm needs parameter optimization during model training and after reducing dimensions to form better clusters.

Ensemble methods have been proposed to overcome the limitations of clustering algorithms when used individually. The main purpose of these methods is to combine algorithms with relative deficiencies with appropriate methods and to apply various dataset features or points with higher accuracy rates in a way that can capture the relationship between them. In the first step, called library generation, the aim is to obtain the set of $m$-many clustering algorithms, $K = \{K_1, K_2,..K_m\}$, where $K$ represents the set of clustering algorithms, and $K_i$ represents each clustering algorithm with different set of parameters, and i=1,...,m. It is revealed that the greater the clustering algorithms in this method focus on various and different features, the greater they can improve the contents of the clusters to be obtained by combining them at the consensus stage~\citep{ensembles_Kuncheva}. Thus, ensembling over a set of differences rather than similar labels will yield more efficient results. The K-means algorithm is generally preferred in the library generation step because of its scalability and low complexity~\citep{ens_Fern, ens_Azimi, ens_Alizadeh, ens_Akbari, ens_Pividori, ens_Yang}. Among the methods used in this step are obtaining results with different starting points or different numbers of clusters with a single algorithm, working on samples with different data points or features, obtaining results on the same data points and features with more than one algorithm or combining several methods. In the consensus step, the $K^{*}$ clustering algorithm, which will be used as the primary clustering method, is obtained from the cluster ensembles within the set of $K$ by integrating results of the clustering with the consensus function. These functions are grouped under four main categories by~\citep{ens_Boongoen} and can be listed as follows: direct, feature-based, pairwise similarity-based, and graph-based approaches.

To compare the performance of the clustering algorithms that are used in the consensus step, in this study, we inspired by a previous work~\citep{hypermeter}, where new clusters are generated by consensus functions from predefined parameter space and compared accordingly. The algorithm procedure is described in Algorithm~\ref{alg:HyperparameterofSamples}. The first strategy obtains the consensus by averaged normalized mutual information (ANMI) maximization. The second strategy consists of two steps. First, the algorithm generates clusters by a consensus function and then finds the labels in the ensemble that share the most information with the consensus step by the maximum normalized mutual information (NMI). It was shown that the optimal hyperparameters found in the second strategy give at least the fitted result as consensus clustering $K^{*}$. Instead of a precise consensus function, in the last step, we applied methods including the cluster-based similarity partitioning algorithm (CSPA)~\citep{ensembles_Ghosh}, the hypergraph partitioning algorithm (HGPA)~\citep{ensembles_Ghosh}, the hybrid bipartite graph formulation (HBGF)~\citep{ens_FernBrodley} and NMF-based (nonnegative matrix factorization) consensus clusterings~\citep{ensembles_DingJordan}) and then selected the one that has the maximum AAMI score (see~\ref{Appendix}). Although we have an ensemble of consensus function results, only the HPGA and NMF methods resulted with the highest AAMI, and among these, the HPGA method was 95\% more voted.

\RestyleAlgo{ruled}
\SetKwComment{Comment}{/* }{ */}
\begin{algorithm}
\caption{Hyperparameter search}\label{alg:HyperparameterofSamples}
\KwData{Data $D~=~\{x_1,x_2,...x_n\}$ to cluster; Set $K$ of ensemble clustering algorithms;
Set $T$ of consensus clustering algorithms}
\KwResult{Matched hyperparameter configuration with the selected consensus clustering}
 \For{$k~\in K$}
{  
Find the clustering results $C(k)$ with the given clustering algorithm $k$ for the investigated dataset\; 
Produce consensus clustering results from the given $T$\;
$t^{AAMI}~=~argmax_{t~\in T} AAMI(C(t),T)$: Find the AAMI result of consensus clustering from produced clustering based on $T$.
}
\Return { $k^{*}~=~argmax_{k~\in K} AMI (C(k),t^{AAMI})$}

\end{algorithm}

\section{Observational Data Analysis} \label{DataAnalysis}
\subsection{Design and Processing of Data Platform}
\begin{figure*}
    \centering 
    \includegraphics[width=15cm]{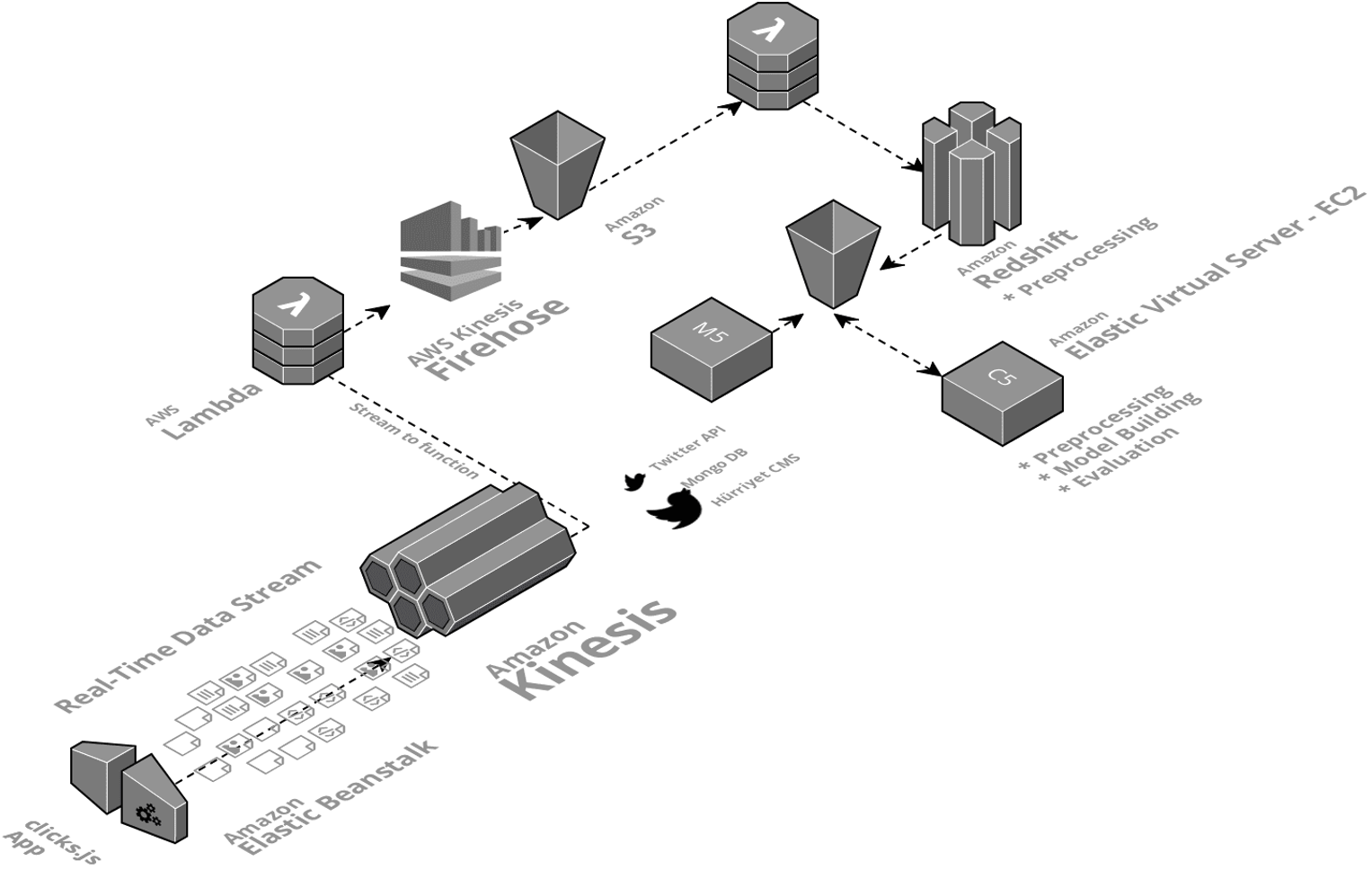} 
    \caption{Clickstream data collection and preprocessing architecture on AWS~\citep{AWS}. The clickstream data, generated by the readers’ navigation actions on the website, and mainly the consecutive HTTP requests from a single IP address, are collected in the S3 buckets following the pipeline (from left to right): H\"{u}rriyet clicks.js application (collects each visit’s action as a request from the website) $\rightarrow$ Elastic Beanstalk (scaled up to 20 virtual machines to meet traffic load) $\rightarrow$ Kinesis Data Streams (scaled up to 30 shards) $\rightarrow$ Lambda (to separate actions by referral types) $\rightarrow$ Kinesis Firehose $\rightarrow$ S3 $\rightarrow$ Lambda $\rightarrow$ Redshift (to provide large-scale query). The datasets are prepared on Redshift by querying and saved to S3 buckets. News content-based datasets (from Mongo DB) and additional sources are also collected on the S3 storage by EC2 virtual machines. Further analyses are processed on the EC2 virtual machines with instance type between C5.2xlarge to c5.4xlarge by reaching S3 buckets.}
    \label{fig:Data_collection_architecture}
\end{figure*}

Clickstream datasets are collected for the research period (April 2019 - May 2020) (Fig.~\ref{fig:Data_collection_architecture}) through web browsers by using Amazon Web Services (AWS; Elastic Beanstalk, Kinesis Data Streams, Kinesis Firehose, API Gateway, Lambda, S3, Redshift, EC2) technologies~\citep{AWS}, which have been actively managed by the company, to enable elastic processing and storage for high-volume data. The collected real-time streaming data are combined and processed on Redshift and then stored on the S3 buckets, which can be accessed through both EC2 virtual machines and local environments. Content-based information collected and combined individually from the company’s servers and AWS.

\textbf{Clickstream Data,}\label{Click-StreamData} stored in separated log files with a total size of 20 TB of sequentially semistructured format data for one year, belong to the visitors’ site history. An identification number (ID) is generated for each anonymized visit on the website, and the data span these visits’ navigation, including actions related to reading, time, browsing platform, content and location. The dataset is aggregated to each visit level such that a single entrance contains all articles read and content-based information included. As stated in previous research~\citep{Makaroglu}, the distribution of total hourly visits referred from~Twitter is similar to the hourly distribution patterns within~Twitter when considering H\"{u}rriyet keywords. Therefore, we deduce that the breakdowns of time-dependent distribution can shed light on the microservice usage patterns; with this inference, timestamps are expanded as $Hour, Week, day of the Week$. The number of distinct articles accessed in a visit is represented as $Visit Depth$ (the total number of contents), rather than embedding as sequential data.~Twitter referrals are filtered from each visit’s referring channel, and the rest of the data are excluded. The category of the articles is represented as categorical names, e.g., travel, magazine, and sports. Only $LeadCategory$ (see Section~\ref{LeadCategory}) is used during the analysis steps. The columnist names of the articles also included in a single dimension.

\textbf{Content-Based Data} spans additional information about articles read by the visitors during the research, including $Posting duration$ (number of days between the date content is published on the website and the date of the visit), $Named Entity Recognition (NER)$ results~\citep{NER}, $Person, Location, Organization$, and the daily  $Content success$ rank. Daily success ranks are calculated by scaling daily articles’ Pageview (PV) scores within their category from 1 (lowest) to 100 (highest). We limit named entities according to the highest TFxIDF (Term Frequency x Inverse Document Frequency) scores within the scope of this study. The missing values in the categorical data types are not imputed; instead, they are represented in the dataset as $unknown:UNK$.

After all the processing, the dataset contains 5,666,125 visits originating from~Twitter generated by 2,261,164 unique visitors, and includes the dimension as follows: $Hour$, $Week$ $Number$, $Week$ $Name$, $Month$, $Location$ $ (City,$~$Country)$, $Device $ $(Mobile, PC)$, $Visit$ $Depth$, $Content$ $Success$, $Posting$ $Duration$, $LeadCategory$, $Columnist$, $Named Entity Recognition$~$(NER\_P (Person),$  ~$NER\_O (Organization),$~$NER\_L (Location))$.

\subsection{Analysis of $LeadCategory$  Distribution} \label{LeadCategory}
\begin{figure*}
    \centering
    \includegraphics[width=\linewidth]{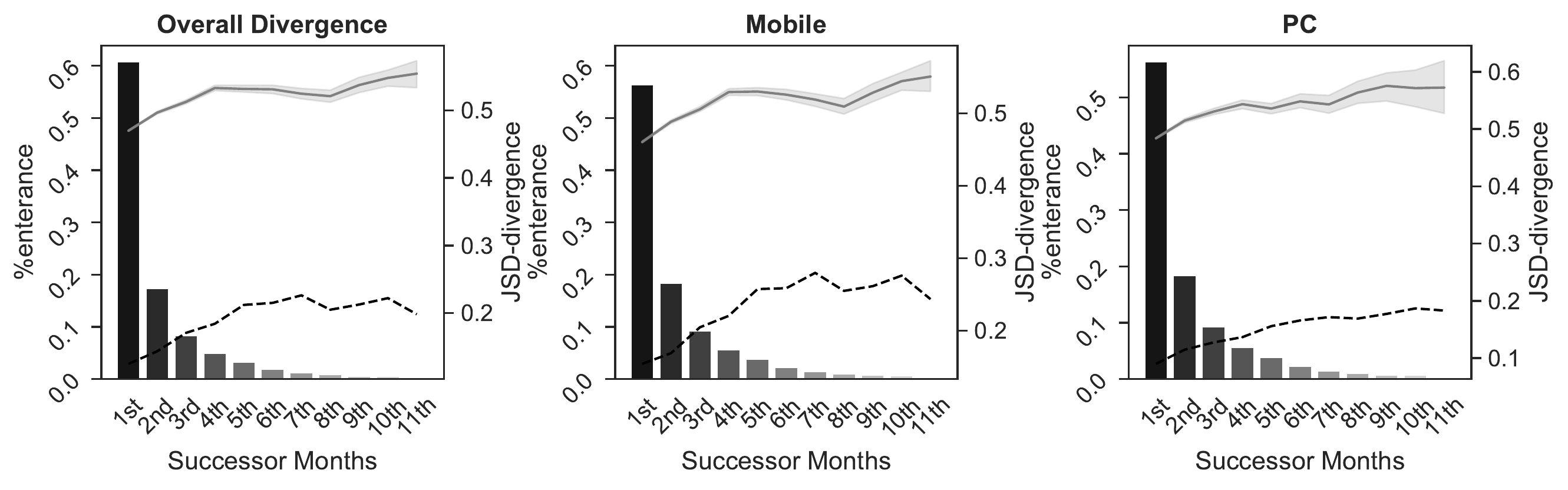} 
    \caption{Comparison of lead category distribution by overall month (dashed black) and device types, including PC and mobile (grey). The X-axis represents the group level of entrances. The left Y-axis represents the rate of entrances within all datasets, and the right Y-axis shows the average of JSD values corresponding to the given groups.}
    \label{fig:Device_breakdown}
\end{figure*}

\begin{figure*}
    \centering
    \includegraphics[width=\linewidth]{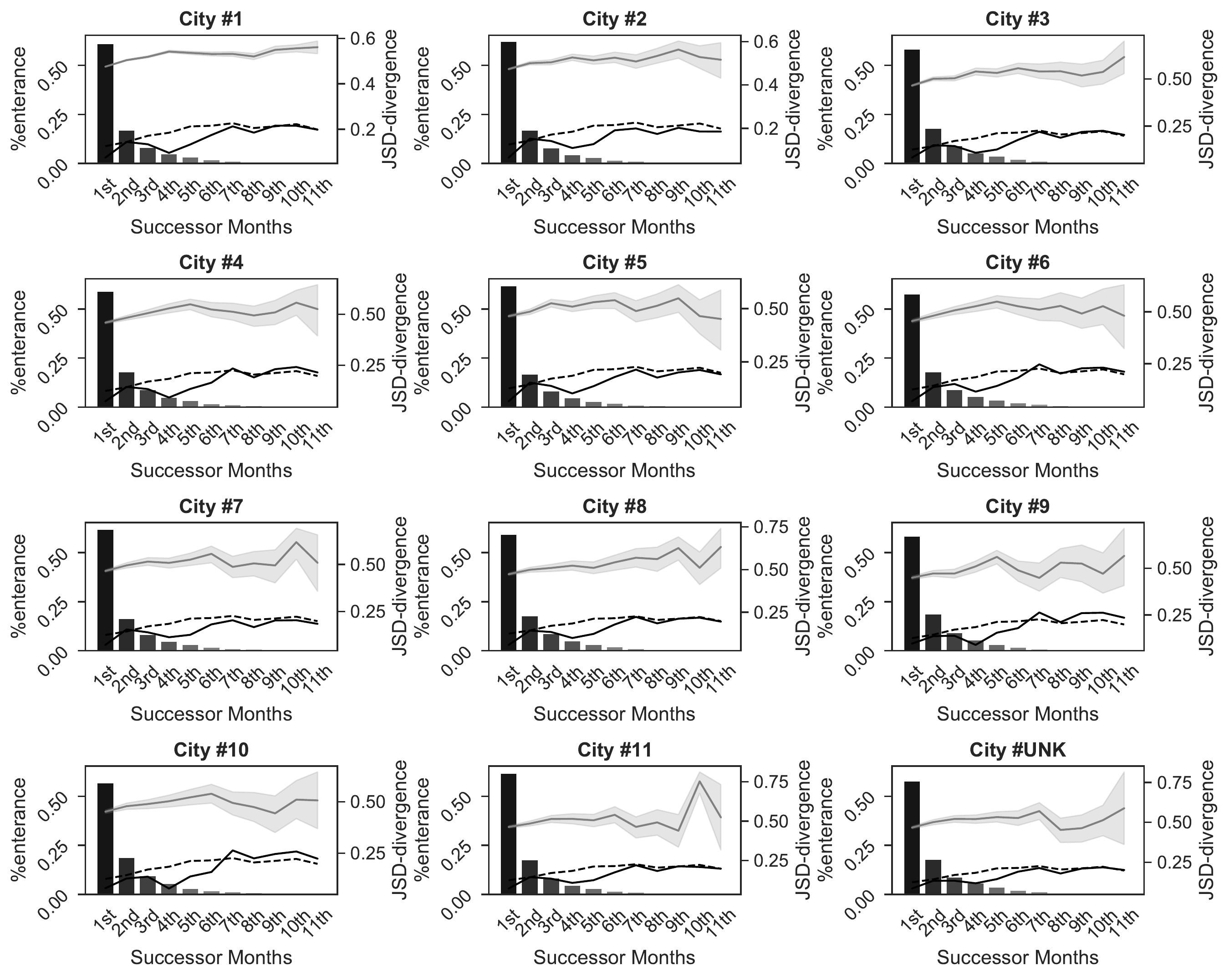} 
    \centering
    \caption{Comparison of city-level lead category distribution by overall month (dashed black), city-level month (black) and individuals (grey). The X-axis represents the group level of entrances. The left Y-axis represents the rate of entrances within all datasets considering the investigated cities, and the right Y-axis shows the average of JSD values corresponding to the groups. The \# numbers represent the different cities, which are positioned in descending order according to the total number of entries they have made during one year.}
    \label{fig:City_caharacteristic_Breakdown}
\end{figure*}

The first article accessed in the referred visit of the streaming data has a prominent role in leading readers from the social media environment and has a directed news consumption with more completion rates~\citep{Bar-Gill}. Early studies showed that the interests in the news topics change over time~\citep{9}. We thus examine the lead categories’ patterns by several breakdowns to show to what extent the distributions change, including browsing platform (to measure the effects of device preferences~\citep{XuForman}), location and compare them with the general reading distributions over months. As described in Section~\ref{Click-StreamData}, each article’s category information is included in the clickstream data with predefined topics $C$~=~$\{c_1, c_2,..,c_n\}$, where $c_i$ represents the topics e.g.~'Current Affairs', 'Sports'. Category-level interests are formulated by dividing total number of clicks, $N$~=~$\{N_1, N_2,..,N_n\}$ of each category topic ($c_i$) with total clicks $N$\textsubscript{total} made by user ($u$) within the investigated month ($m$) such that the outputs are given as the following distribution vectors:

\begin{equation}
	{T(u,m)~=~}\Bigg(\dfrac{N_1}{N_{total}},\dfrac{N_2}{C_{total}},\dots,\dfrac{N_n}{N_{total}}\Bigg), \quad {N_{total}~=~}\sum_{i}^{m}{N_i},
	\label{art_dist}
\end{equation}	
where $N_{i}$  is the total number of pageviews that the user made on the referred articles that are classified as the specific category, ($c_i$), and $N_{total}$  is the total number of clicks that the user made for the given month. These distributions are found by considering individual anonymized user~$ID$s' as well as the~$Overall$ group of monthly readers. However, the referral~$ID$s vanished quickly for each breakdown (Figure~\ref{fig:Device_breakdown}). A possible explanation for the observed differences in the rates of enjoyment is that the readers can systematically clear the ~$ID$s. To rule out this possibility, rather than tracking the individuals, we grouped them within discrete time windows, such that in case the referral~$ID$ shows up in one month, its group is defined as $1^{st}$, if it is in two months, then the group is defined as $2^{nd}$, until the $11^{th}$ entrance in 1 year. Relatedly, all monthly user distributions are computed by dividing one year into 12 months and comparing divergence by grouping successor months as $1^{st}$,$2^{nd}$,..$11^{th}$. The Jensen–Shannon Divergence (JSD) metric (see Appendix~\ref{Appendix}) is used to measure similarity with the given distributions. We obtain category-based characteristics by averaging each group’s JSD value. Figure~\ref{fig:Device_breakdown} and \ref{fig:City_caharacteristic_Breakdown} depict that the lead category interests are differentiated across the grouped months and show that the overall month distribution is more constant than that of individuals in all three breakdowns, including Mobile, PC(Desktop), and City.

\section{Methodology} \label{Methodology} 
This study utilizes an ensemble clustering evaluation method for patterns based on features extracted from reader-ship browsing and content-based data. In Section~\ref{EmbeddingHighDimensions}, we explore different combinations of two embeddings as input representations (i.e., UMAP, N2d). In Section~\ref{SamplingApproach}, we discuss the sampling approaches and evaluation of the selection of the minimum sample size. In Section~\ref{Proposed_Approach_to_Ensemble_Clusterng}, we experimentally investigate ensemble clustering methodologies and evaluate the optimal clustering methods. In Section~\ref{Evaluation}, we discuss the results of the conducted experiments for the interpretation of reader behaviour stability that we obtained.

\subsection{Embedding High Dimensions} \label{EmbeddingHighDimensions}
When the correlation of categorical and numerical features before proceeding to the clustering step is below a specific value, the success in this step is affected, as it can eliminate the redundant information during clustering. When we examined the correlation in all data types (numerical, categorical, ordinal), we observed that it changes within the limits of 0.2 to 0.75 over the months. Therefore, before proceeding to the clustering step, it is confirmed that none of the correlations found surpass the maximum cut-off of 0.75. One hot encoding is applied for categorical data, and power transformation is applied for numeric data types. To make the numerical values more Gaussian-like, in this step, Yeo-Johnson’s transform~\citep{yeo_johnson} is applied, which supports both positive and negative values and makes the data zero-mean while minimizing skewness. After the encoding step, we experiment with two embedding representations named uniform manifold approximation and projection (UMAP)~\citep{umap} and N2d~\citep{N2D}. 

UMAP uses a graph-based approach to project the high-dimensional space to the lower dimension while preserving the global structure. The L\textsubscript{2} norm for numeric values and the Dice distance for categorical data types were used as distance criteria. As stated in the paper, with this method, we can generate fuzzy simplicial sets independently. UMAP supports intersecting two graphs and then embedding them as a consensus graph in low-dimensional space to create a single fuzzy simplicial set. The intersected graph neither has mostly dense areas as in categorical representation nor is weakly connected, as in the numerical graph. Instead, it has the middle range of all connectivity placed by the probability of edge existence. To achieve this unity, we used percentages of categorical columns as weight parameters. 

N2d lowers the dimensions by creating autoencoders with a trained neural network. After having a dense layer of created autoencoders, this method adds manifold learning to increase clusterability performance by preserving local structures and capturing global structures. The article~\citep{N2D} states that if the total number of clusters in the dataset is known beforehand, the size in the UMAP stage should be defined accordingly. In this approach, we compared the 10 dimensions used in the original article to represent 10 known clusters and the 2-dimensional UMAP results after learning the autoencoders that we used in the first method as the final representation dimension of the dataset. Since our target is having clustering $C~=~F_C(F_M(F_A))$, with a better represented embedding space applied on dataset $D$ with $n$ data points in $d$ dimensions $D~=~\{x_{1},x_{2},...x_{n}\}\subseteq\ R^{d}$ in this approach, we are searching for the most applicable clustering algorithm ($F_C$), after having a manifold learner ($F_M$) of the autoencoder ($F_A$). 

\subsection{Sampling Approach} \label{SamplingApproach} 
Most of the clustering algorithms’ complexity is related to the number of sample data points. Therefore, it is both efficient and necessary to work on small-sized samples that can accurately represent the data, especially when working with large datasets. For this purpose, based on the results we obtained in Section~\ref{LeadCategory}, where we observed that the interests of the users change over time by the news category, we select samples from each month to keep correlations among news interests as well as to include current news events that took place in that month and correlated with each other. Stratified sampling is an approved sampling strategy that captures the main characteristics from the given dataset, is profoundly applied in large-scale datasets, and gives high accuracy when sampled from the features~\citep{Stratified}. We compared different sampling strategies within the same months to observe the minimum sample size that we could proceed with, including monthly behaviours. Therefore, on the samples that we selected both at random and stratified by news categories from each month, the K-means algorithm is used to evaluate the internal validation of the clusters, which is scalable and can return fast results. Silhouette coefficient (SI)~\citep{Silhouette}, Calinski Harabasz index (CHI)~\citep{calinski},  Davies Bouldin scores (DB) and Dunn index-type (DI)~\citep{Dunn} measurements were compared by evaluating the number of clusters between 2 and 30 on four random and four stratified samples that we obtained from 5000 to 30000 in increments of 1000 data points. To evaluate the clustering tendency of each sample, the Hopkins test~\citep{hopkins} is applied to the sampled datasets, which gave values close to 1 for highly clusterable data. According to the result of this test, if the values are close to 0.5, it is concluded that the data are uniformly distributed, and therefore, clusterability is not possible. This value is found to be above 0.95 for all sample sizes for the given months, and as the sample size is increased, its clusterability also increases. Although the results of UMAP give lower values than the N2d approaches, especially for certain months, its clusterability is within the range of acceptable values that remain above 0.95.

\subsubsection{Evaluation of sampling by internal metric validations} 
To evaluate the approach that the subsets of sampling strategies are related to each other, the similarity of each validation with the other results in its group is investigated. The adjusted mutual information (AMI)~\citep{AMI} (See Appendix~\ref{Appendix}) score is computed during this evaluation because it could measure the success of the groups’ association, regardless of the order of the labels and the sample size. An overview of this process can be found in Algorithm~\ref{mthautorefname:AAMIofSamples}. First, the metric scores correspond to the $k$ values calculated concerning the selected sample size range in each month’s random and stratified sampling strategies. After that, they were binned with the Freedman-Diaconis rule~\citep{FD_rule}. While binning, each sampling strategy is evaluated between its minimum and maximum values in its primary group as random and stratified, so the results of different approaches are not suppressed by each other. After the metric results are binned and labelled, AMI scores are calculated. To compare the validation information that a sample size shares with the sample group, the Averaged Adjusted Mutual Information (AAMI), which has proven success in ensemble methods~\citep{ensembles_Ghosh, hypermeter}, is computed.

\RestyleAlgo{ruled}
\SetKwComment{Comment}{/* }{ */}
\begin{algorithm}
\caption{Search of sample size} \label{mthautorefname:AAMIofSamples}
\KwData{Set $K$ of cluster sizes; Set $M$ of internal validation metrics; data $D~=~\{x_1,x_2,...x_n\}$ to cluster; investigated sample range $S$}
\KwResult{AAMI of each sample size for the evaluated metric}
 \ForAll{$s~\in S$, $m~\in M$, $k~\in K$}
{  
 Find internal validation score for each metric $m$ in the set of $M$ correspond to the cluster size of $k$ in the set of $K$, for selected s in the set of $S$ \;
 Scale the results between [0,1] within the given internal validation metric $m$ \;
 Bin the scaled results by FD rule within the given $m$ \; 
 }
 \Return {$Bin_{FD}(Scale(M(s,m,k)))$ vectors for each range of sample size}\;  
\ForAll{$s_i,s_j~\in S_{i \neq j}$}
{ $AMI(s_i,s_j)$\; 
Sum the AMI values for each sample $s_i$, that are calculated between each vectors of size $|K|$ for the given $m$ for each sample\;
Divide the Sum by the size of total range $|S|$ for each $s_i$ to find average adjusted mutual information of the given sample $s_i$.
}
\Return  {$AAMI(s_i)$} 
\end{algorithm}

\begin{figure*}
    \centering
    \includegraphics[width=17cm]{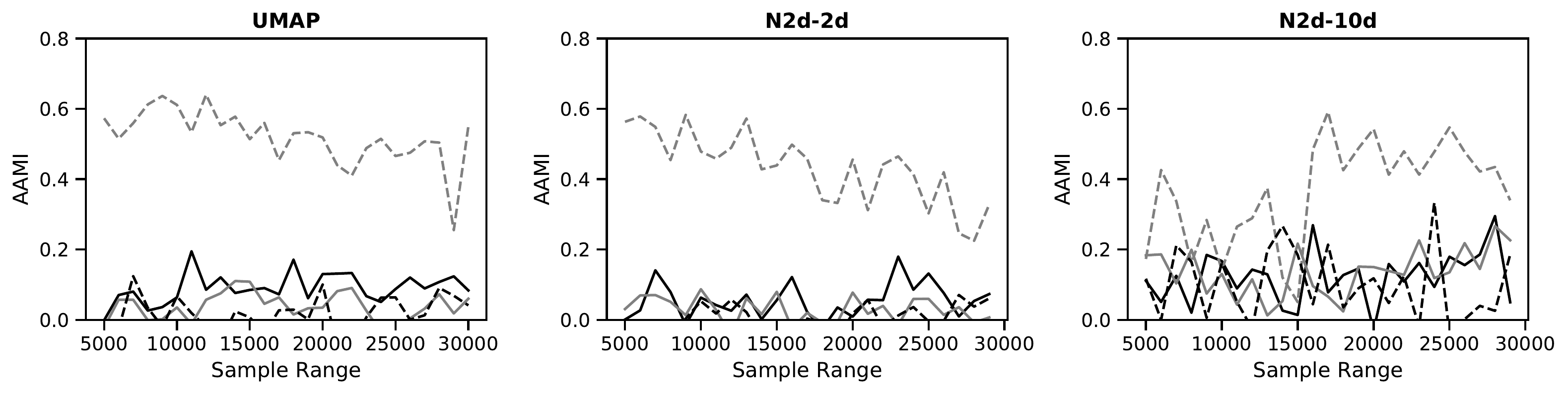}
    \caption{UMAP-2d, N2d-2d, and N2d-10d obtain AAMI test results according to sample sizes. The colours in the graph lines represent applied internal validation metrics that are SI, DB, DI, and CH given in black, grey, dashed black, and dashed grey, respectively. One of the stratified samplings of November is selected for representation purposes. The X-axis represents the sample sizes increasing from 5000 to 30000. The y-axis of the grid picture represents the AAMI values.}
    \label{fig:ANMI_scores}
\end{figure*}

One of the stratified sampling results from November is shown in Figure~\ref{fig:ANMI_scores}. Except for CHI, the AAMI values remained relatively stable even when the sample size was increased, which we expected to observe to set the minimum sample size. Even though we used internal validity scores to determine the sample size, considering only the distance of the data points intra- or inter-clusters may cause nonconvex geometries not to be detected, or this approach might not be meaningful for the cluster centres that are embedded in nonlinear sub manifolds~\citep{hypermeter}. The fact that any label cannot be known beforehand due to unsupervised learning problems confirms that we cannot use the external validity index in this problem where we aim to cluster user behaviours. For these reasons, we performed a knowledge-based assessment to confirm the stability of the resulting cluster in Section~\ref{ClusterStability}. 

\subsection{Methodology of Ensemble Clustering} \label{Proposed_Approach_to_Ensemble_Clusterng} 
We applied a method from a previous study~\citep{hypermeter} and referred to Strategy 1 and our approach to Strategy 2, which is shown in the model architecture in Figure~\ref{fig:model_architecture}. In the library step, we applied eight algorithms combined with the following different parameters: K-means, Agglomerative, Spectral, Birch, GaussianMixture, HDBSCAN, DBSCAN, and Affinity Propagation. Therefore, the complexity of the local step is the sum of the complexity of the algorithms (see Table~\ref{tab:table_alg}). Once the Library step’s clustering results are produced, their resulting clusters are found by applying four different consensus functions at the consensus step. The final consensus function is selected as the one that gives the maximum AAMI result among four different functions. For the final step of Strategy 1, the best match is selected by the maximum AMI result given between the consensus function and the hyperparameter set of ensembles.

  \begin{table}[hbt!]
   \centering
     \begin{tabular*}{\tblwidth}{lp{7.5em}l}
     \toprule
     \multicolumn{1}{p{8.165em}}{Algorithm} & \multicolumn{1}{c}{Time Complexity  } & \multicolumn{1}{p{7.835em}}{Category} \\
     \midrule
     \midrule
     K-means    & $\mathcal{O}(kNT)$         & Partitioning \\
     Agglomerative  & \multicolumn{1}{l}{ $\mathcal{O}(N^2)$} & Hierarchial \\
     Spectral  & \multicolumn{1}{l}{$\mathcal{O}(N^3)$ } & Model-Based \\
     HDBSCAN   & \multicolumn{1}{l}{$\mathcal{O}(N\log N)$} & Density-Based \\
     DBSCAN    & \multicolumn{1}{l}{$\mathcal{O}(N\log N)$ | $\mathcal{O}(N^2)$ } & Density-Based \\
     Birch     & \multicolumn{1}{l}{$\mathcal{O}(N)$} & Hierarchial \\
     GaussianMixture & \multicolumn{1}{l}{$\mathcal{O}(N'2KD)$} & Distribution \\
     AffinityPropagation & $\mathcal{O}(N^2 \log N )$         & Model-Based \\
     \bottomrule
     \bottomrule
     \end{tabular*}%
   \label{table_alg}
\caption{\label{tab:table_alg}N:Number of samples, k:number of clusters, T:number of iterations, D:dimension }
 \end{table}%

\begin{figure*}
    \centering
    \captionsetup{justification=centering,margin=2cm}
    \includegraphics[width=17cm, height=19cm]{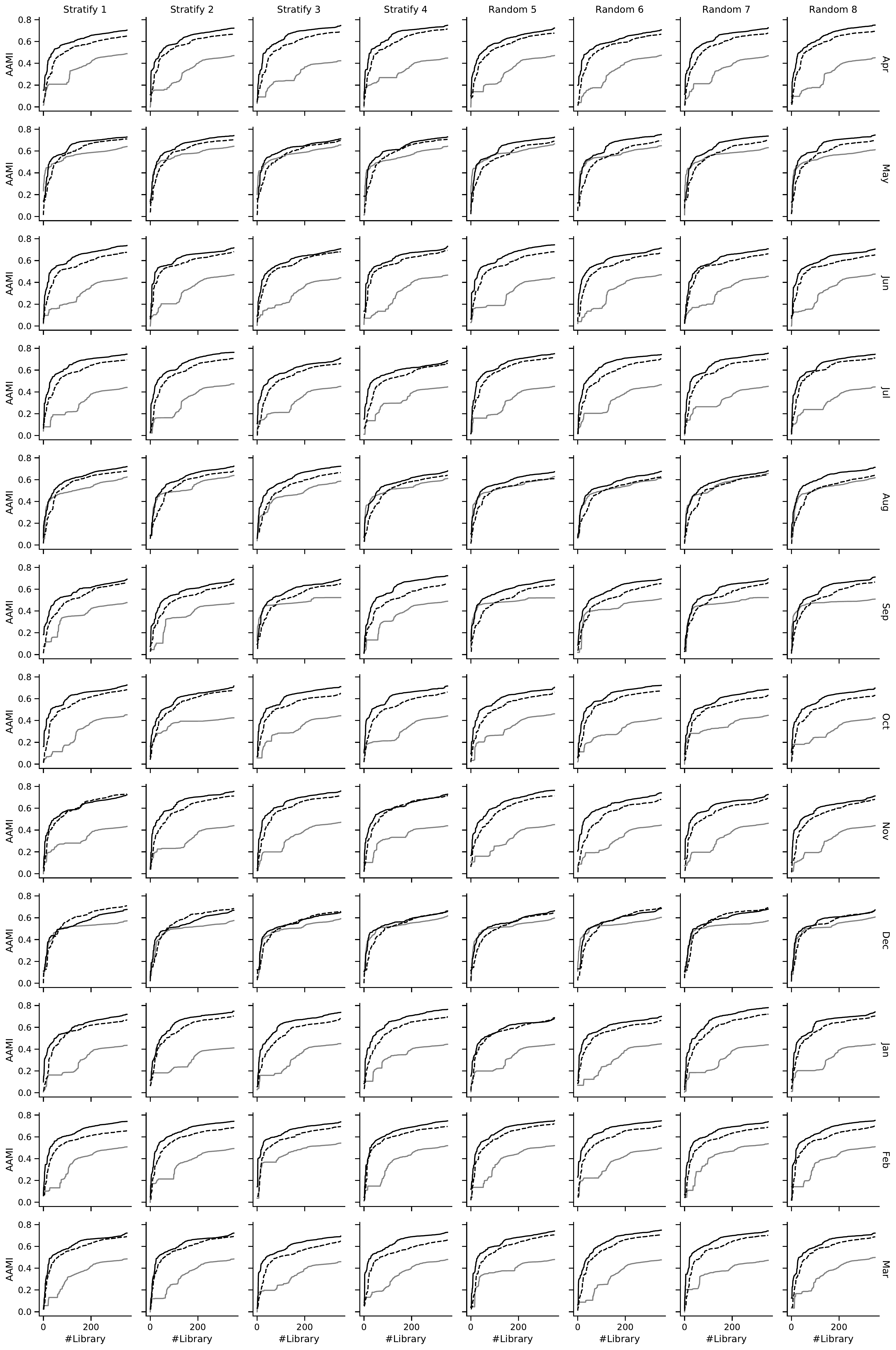}
    \caption{AAMI results are obtained after the decreased ensemble approach. The colours in the graph lines represent applied embeddings that are UMAP, 10-dimensional N2d, and 2-dimensional N2d given in grey, dashed black, and black, respectively. The rows represent months, and the columns represent the sampling strategy with index numbers. Each chart includes the number of clustering configurations on the X-axis, while the Y-axis shows the total selections.}
    \label{fig:ANMI_distribution_for_decreased_ensembles}
\end{figure*}

 For the first step of Strategy 2, we consider clustering to be obtained with the consensus function. As explained in Section~\ref{Literature}, the higher the variance, the more accurate the results that can be extracted. To this end, a new ensemble set is obtained by sorting AAMI values from low to high and then eliminating those above a particular threshold value to reduce redundancy and complexity. Thus, clusters that produce similar labels are removed from the ensemble, and clustering is obtained with the consensus function among the remaining clusters. As shown in Figure~\ref{fig:ANMI_distribution_for_decreased_ensembles}, when the AAMI scores of the relevant approach are sorted and observed according to the number of algorithms in the entire ensembles, it is revealed that both approaches have different threshold values. Regarding N2d, these values show similar patterns in 10- or 2-dimensional representations, which is stabilized earlier than nearly half the value for UMAP. Considering that the AAMI results of different approaches have different distributions, threshold values of 100 and 200 are set for N2d and UMAP, respectively. After that, all steps in Strategy 1 are applied to this new set. As a result, labels are obtained with six different clustering approaches for each dataset.
 
 \subsection{Evaluation} \label{Evaluation}
 \subsubsection{Clustering Results} 
 The primary purpose of clustering validation is to discriminate among reading behaviour groups, which would help newsrooms recommend specific news for those in the given clusters. Since we have 6 different strategies with all consensus clustering labels, there are 356 possible clustering solutions in total. Therefore, the evaluation of the resulting clustering is nontrivial for this study since there is no ground truth available in the data and the knowledge-based interpretations are subjective. Consequently, in the first step, we aimed to reach the best evaluation results among the clustering approaches. To identify successful strategies, internal validation metrics (SI, CH, DB, S-Dbw~\citep{Sdbw_Kim, Sdbw_Tong, Sdbw_Halkidi} (see Appendix~\ref{Appendix}) are compared based on the results of the clustering algorithms obtained for the month, embedding and sampled dataset. With this, we could observe both compactness/separation and arbitrary shape density validation of the clusters. As stated in a previous study~\citep{Handl}, internal validation metrics must be selected by the applied clustering algorithms since both of them consider the underlying structure of the given dataset. Therefore, the metrics, especially when the ground truth is not known, are crucial for validating the resulting clusters. As reviewed in the study~\citep{evaluation_review},  metrics also come from certain assumptions, so there is a trade-off in the choice of these evaluations. In the given review study, cluster validation evaluated four main categories: Cluster (shape (spherical/convex, nonstandard geometry, arbitrary) and numbers (can be attributed to the high number of clusters)), Handling Noise (ability to deal with outliers without preprocessing) and Computational cost/complexity. On the other hand, capability with subcluster identification and skew distribution were also evaluated by another study~\citep{Clustering_Validation_Measures} (See Table~\ref{tab:table_metric}). 

\newcommand\addcaptionbelow[1]{{%
  \let\caption\captionbelow
  \caption*{#1}%
}}

 \begin{table}[hbt!]
   \centering
   \caption{\label{tab:table_metric}Evaluated metrics' comparison}
     \begin{tabular*}{\tblwidth}{lllllll}
     \toprule   
     \multicolumn{1}{p{1.585em}}{Metric} & \multicolumn{1}{p{2.9em}}{\begin{tabular}{@{}c@{}} \#Cluster \\ Bias \end{tabular}} & \multicolumn{1}{p{2.9em}}{\begin{tabular}{@{}c@{}}Handling \\ Noise\end{tabular}} & \multicolumn{1}{p{2.95em}}{\begin{tabular}{@{}c@{}}Sub \\ Clusters\end{tabular}} & \multicolumn{1}{p{2.75em}}{\begin{tabular}{@{}c@{}}Skew \\ Disribution\end{tabular} } & \multicolumn{1}{p{2.5em}}{\begin{tabular}{@{}c@{}}Optimal \\ Value\end{tabular}} \\
     \midrule
     \midrule
     SI         & No        &  \multicolumn{1}{p{2.5em}} {Depends on\newline{}Noise(\%) } & No        & Yes       & Max \\
     CHI       & No        & No        & Yes       & No        & Max \\
     DB        & No        & Yes       & No        & Yes       & Min \\
     S-Dbw \\ (Halkidi) & Yes       & Yes       & Yes       & Yes       & Min \\
     S-Dbw \\ (Kim) & Yes       & Yes       & Yes       & Yes       & Min \\
     S-Dbw \\ (Tong)& No        & Yes       & Yes       & Yes       & Min \\
     \midrule
     \multicolumn{6}{p{25.59em}}{S-Dbw (Halkidi)~\citep{Sdbw_Halkidi}, 
     S-Dbw (Kim) \citep{Sdbw_Kim}, 
     S-Dbw (Tong) \citep{Sdbw_Tong}} \\
     \bottomrule
     \bottomrule
     \end{tabular*}%

 \end{table}%

\begin{figure*}
    \centering
    \includegraphics[width=18cm, height=10cm]{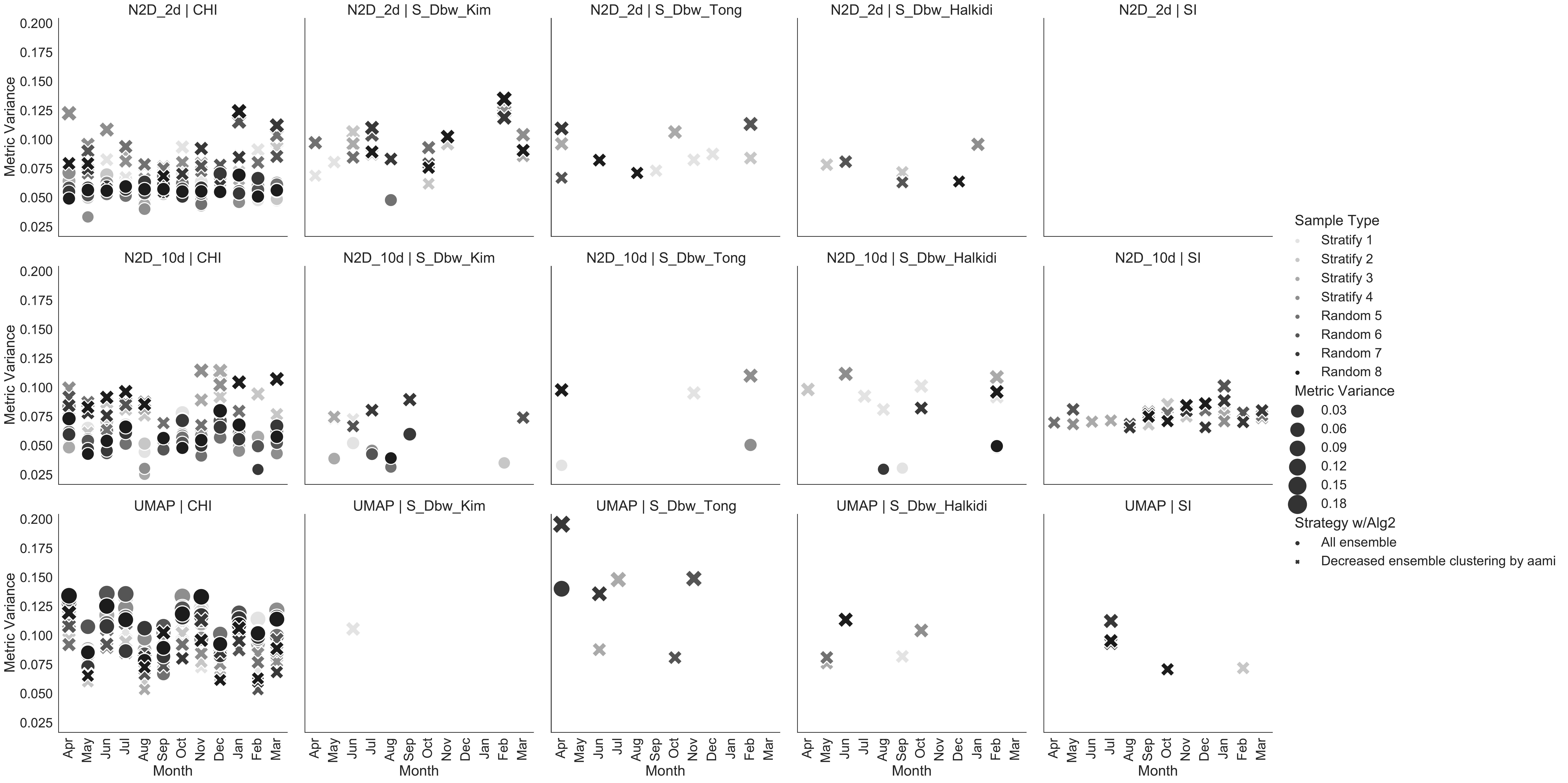}
    \caption{Metric selection by individual datasets sampled from each month and applying different strategies to embeddings. The cross sign "x" represents the optimal library metric results according to the decreased ensemble library, and sphere "o" represents the entire ensemble set.}
    \label{fig:Individualized_metrics}
\end{figure*}

\begin{figure*}
    \centering
    \includegraphics[width=\textwidth, height=7cm]{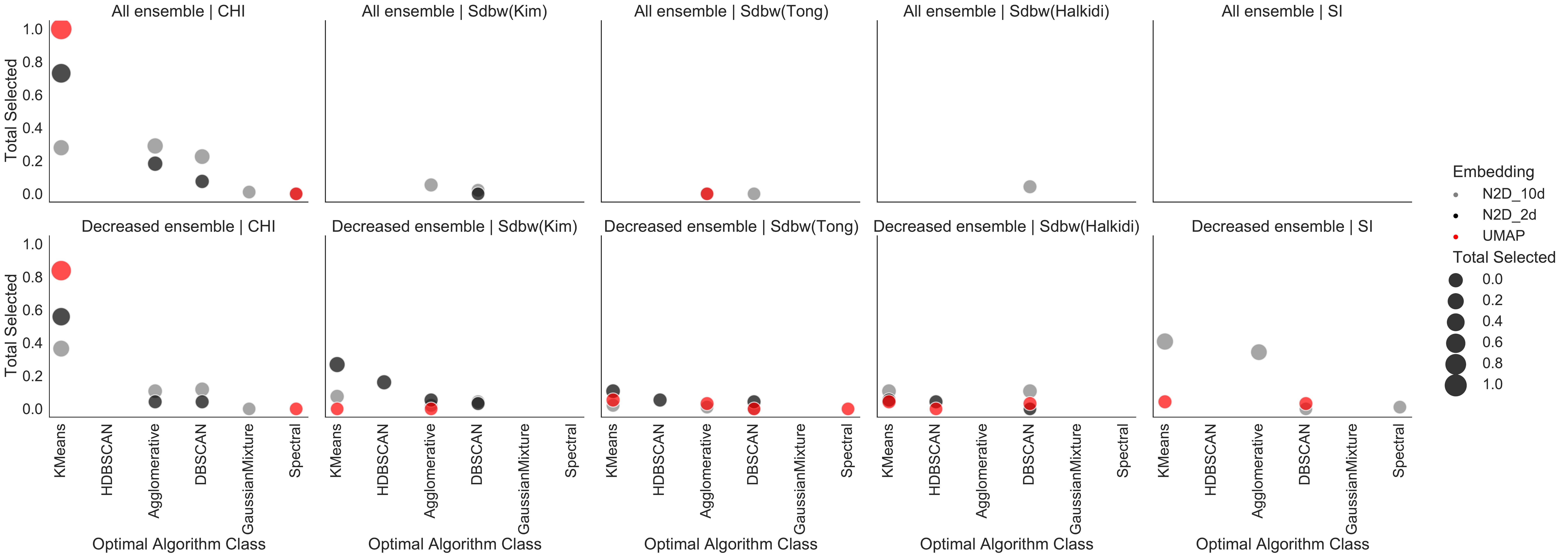}
    \caption{Optimal metrics obtained according to the maximum variance selection and algorithms that take their values at the maximum limits of these metrics. While the columns show the selected metrics, each strategy groups are represented in the rows. Embedding breakdowns are discriminated by the colors, where gray, black and red represent N2d-10d, N2d-2d, and UMAP, respectively. The area size of the circles scales the total number of selections.}
    \label{fig:algorithm_groups_by_metrics}
\end{figure*}

Therefore, we evaluated cluster performances in a format that included various validation perspectives, as different characteristics of clusters could be focused on.
Internal validation values are used as a measure to determine the optimality as the quality of clusters, which gives successful results in a previous study on identifying behaviours~\citep{Fernandez_clustering}. Following this purpose, internal validation metrics of the clusters formed by the algorithms within the ensemble groups created separately by Strategies 1 and 2 are evaluated. Considering the optimal value of the metrics at the minimum and maximum limits, all scores are unit-based scaled between [0,1] within the evaluated dataset. Then, the successful metric was determined according to the one with the highest variance among these lists. As shown in Fig~\ref{fig:Individualized_metrics}, although there are standard metrics, the majority of the selection belongs to CHI(\%76) among all, and the rest of the top selections are S-Dbw~\citep{Sdbw_Kim}($\%$8), SI($\%$7), S-Dbw~\citep{Sdbw_Halkidi}($\%$5) and S-Dbw~\citep{Sdbw_Tong}($\%$4). The algorithm classes that take the best values in the limits of the metrics with the maximum variance are shown in Fig~\ref{fig:algorithm_groups_by_metrics}. According to the results, while the most successful algorithm for CHI is K-means, HDBSCAN did not give any results with CHI, which obtains the best validation scores with S-Dbw. Therefore, this result also supports the internal validation approach of S-Dbw, which considers the density area in the selected cluster.

\subsubsection{Selection of Successful Strategy} 
To select the most successful embedding and strategy, we utilize the closeness of the strategies and embedding approaches to the optimal metric scores. Although the CHI metric is found to give a generalizable result, it has been observed that different metrics can give optimal results in some months due to the differences in the distributions of the monthly dataset samples. For this reason, the metric values obtained with the strategies and approaches are evaluated with the filtering approach described in Table~\ref{tab:filteringapproach}. 

\begin{enumerate}
\item  First, the algorithms, which are given by the metrics obtained according to the maximum variance calculation, are extracted (Optimal/Metric). Then, considering the optimal value of the metrics at the minimum and maximum limits for each breakdown, the top 5 results with the best score are obtained and ranked between 1 (the best) and 5 (the least). Thus, optimal metrics giving the best five scores for Strategy 1 and Strategy 2 and algorithms providing these metrics (Optimal/Algorithm) are prepared.
\item The closeness of the applied strategies’ metric scores is calculated by taking the differences of the optimal metric values, which can be selected differently from CHI according to the datasets, and the corresponding metric values of the applied strategies on that dataset. The value differences are then scaled within the observed metrics between [0,1]; 0 represents the closest to the optimal value, while 1 represents the farthest (Evaluation/Closeness). 1 is the closest to the optimal value.
\item The top 5 optimal algorithms filtered in the first step are compared with the algorithms obtained with the strategies, and their performances are weighted with specific coefficients. These coefficients are determined so that if an exact match with the algorithm (including all hyperparameters) occurs (Evaluation/Exact Match), the optimal algorithm’s rank is (1-5), and if not, the minimum value (6) is placed (Evaluation/Rank Weighted). We evaluated the closeness of the clusters formed by the consensus function to the optimal algorithm with the ranking value obtained by the hyperparameter match method, since there can be no match with any optimal algorithm in the consensus step. Because the clusters are obtained as a result of consensus, the algorithm with the highest similarity to the library is found by the hyperparameter match method. Therefore, this similarity gives us the similarity of the consensus result to the optimal values (Evaluation/Rank Weighted, in red). However, we observed that the internal validation scores of the optimal algorithms on the samples give more successful results than the clusters obtained by consensus.
\item The scaled closeness values, which are obtained in the second step, are multiplied by the coefficients obtained in the third step (Evaluation/Rank Weighted * Evaluation/Closeness~=~Evaluation/Weights). Then, based on each sample, the results are ranked between 1 (the best) and 6 (the farthest) (Evaluation/Rank). Therefore, considering the entire dataset, the embedding approach with the highest number of times ranked 1 is the closest to the optimal value. 
\end{enumerate}

The successful strategy is found by following the applied embedding considering the ranking criterion given in the Evaluation/Rank column (Table~\ref{tab:filteringapproach}). Thus, the results are ranked from minimum values to give the closest result to the optimal value and the maximum values to give the farthest result. At the same time, the performances of strategies that can achieve one-to-one matches were also weighted. In this step, we also discriminate the rankings as stratified and randomized to compare the sampling types.

\begin{sidewaystable}[h!p hbt!]
\caption{Clusters with the highest reoccurrence counts}
\scriptsize
\setlength\tabcolsep{3pt}  
\setlength\extrarowheight{-5pt}

    \begin{tabular}{p{5.335em}lllllllllllllll} 
 \multicolumn{16}{c}{} \\
 \midrule
 \multicolumn{4}{|c|}{\textbf{Strategy, Dataset, Approach}} & \multicolumn{4}{c|}{\textbf{Optimal}}             & \multicolumn{3}{c|}{\textbf{Selected}} & \multicolumn{5}{c|}{\textbf{Evaluation}} \\
 \midrule
 \multicolumn{1}{|p{5.335em}|}{\textit{Month /Sample}} & \multicolumn{1}{p{5.75em}|}{\textit{Embedding}} & \multicolumn{1}{p{4.5em}|}{\textit{Strategy w/Alg}} & \multicolumn{1}{p{6.085em}|}{\textit{Strategy}} & \multicolumn{1}{p{3.915em}|}{\textit{Class}} & \multicolumn{1}{p{6em}|}{\textit{Hyper-parameters}} & \multicolumn{1}{p{5.085em}|}{\textit{Metric - Value}} & \multicolumn{1}{p{4.75em}|}{\textit{\#Clusters}} & \multicolumn{1}{p{8.165em}|}{\textit{Hyper-parameters}} & \multicolumn{1}{c|}{\textit{Class}} & \multicolumn{1}{c|}{\textit{\#Clusters}} & \multicolumn{1}{p{3.585em}|}{\textit{Exact Match}} & \multicolumn{1}{p{4.915em}}{\textit{Closeness}} & \multicolumn{1}{p{5.085em}|}{\textit{Rank Wieghted}} & \multicolumn{1}{p{4.335em}|}{\textit{Weights}} & \multicolumn{1}{p{2.835em}|}{\textit{Rank}} \\
 \midrule
 \multicolumn{1}{|c|}{\multirow{18}[12]{*}{May /Random 6}} & \multicolumn{1}{c|}{\multirow{6}[4]{*}{N2d\_10d}} & \multicolumn{1}{c|}{\multirow{3}[2]{*}{DE}} & \multicolumn{1}{c|}{CC- Stg 2} & \multicolumn{1}{c}{\multirow{3}[2]{*}{Agg}} & \multicolumn{1}{c}{\multirow{3}[2]{*}{L2\_Avg\_2}} & \multicolumn{1}{c}{\multirow{3}[2]{*}{SI - 0,662}} & \multicolumn{1}{c|}{\multirow{3}[2]{*}{2}} & \multicolumn{1}{c}{hgpa} & \multicolumn{1}{c}{HGPA} & \multicolumn{1}{c|}{163} & \multicolumn{1}{c}{0} & \multicolumn{1}{c}{0.56} & \multicolumn{1}{c}{6} & \multicolumn{1}{c}{3.34} & \multicolumn{1}{c|}{6} \\
 \multicolumn{1}{|c|}{} & \multicolumn{1}{c|}{} & \multicolumn{1}{c|}{} & \multicolumn{1}{c|}{HM - Stg 2} &            &            &            & \multicolumn{1}{c|}{} & \multicolumn{1}{c}{k~=~4} & \multicolumn{1}{c}{K-means} & \multicolumn{1}{c|}{4} & \multicolumn{1}{c}{0} & \multicolumn{1}{c}{0.32} & \multicolumn{1}{c}{6} & \multicolumn{1}{c}{1.93} & \multicolumn{1}{c|}{5} \\
 \multicolumn{1}{|c|}{} & \multicolumn{1}{c|}{} & \multicolumn{1}{c|}{} & \multicolumn{1}{c|}{ANMI - Stg 2} &            &            &            & \multicolumn{1}{c|}{} & \multicolumn{1}{c}{Euc\_comp\_14} & \multicolumn{1}{c}{Agg} & \multicolumn{1}{c|}{14} & \multicolumn{1}{c}{0} & \multicolumn{1}{c}{0.09} & \multicolumn{1}{c}{6} & \multicolumn{1}{c}{0.47} & \multicolumn{1}{c|}{\cellcolor[rgb]{ .851,  .851,  .851}\textbf{1}} \\
\cmidrule{3-16} \multicolumn{1}{|c|}{} & \multicolumn{1}{c|}{} & \multicolumn{1}{c|}{\multirow{3}[2]{*}{AE}} & \multicolumn{1}{c|}{CC- Stg 1} & \multicolumn{1}{c}{\multirow{3}[2]{*}{K-means}} & \multicolumn{1}{c}{\multirow{3}[2]{*}{k~=~30}} & \multicolumn{1}{c}{\multirow{3}[2]{*}{CHI - 8459,528}} & \multicolumn{1}{c|}{\multirow{3}[2]{*}{30}} & \multicolumn{1}{c}{hgpa} & \multicolumn{1}{c}{HGPA} & \multicolumn{1}{c|}{290} & \multicolumn{1}{c}{0} & \multicolumn{1}{c}{0.23} & \multicolumn{1}{c}{6} & \multicolumn{1}{c}{1.40} & \multicolumn{1}{c|}{4} \\
 \multicolumn{1}{|c|}{} & \multicolumn{1}{c|}{} & \multicolumn{1}{c|}{} & \multicolumn{1}{c|}{HM - Stg 1} &            &            &            & \multicolumn{1}{c|}{} & \multicolumn{1}{c}{L1\_15\_0.8} & \multicolumn{1}{c}{DBSCAN} & \multicolumn{1}{c|}{68} & \multicolumn{1}{c}{0} & \multicolumn{1}{c}{0.21} & \multicolumn{1}{c}{6} & \multicolumn{1}{c}{1.28} & \multicolumn{1}{c|}{3} \\
 \multicolumn{1}{|c|}{} & \multicolumn{1}{c|}{} & \multicolumn{1}{c|}{} & \multicolumn{1}{c|}{ANMI - Stg 1} &            &            &            & \multicolumn{1}{c|}{} & \multicolumn{1}{c}{Euc\_avg\_26} & \multicolumn{1}{c}{Agg} & \multicolumn{1}{c|}{26} & \multicolumn{1}{c}{0} & \multicolumn{1}{c}{0.09} & \multicolumn{1}{c}{6} & \multicolumn{1}{c}{0.51} & \multicolumn{1}{c|}{2} \\
\cmidrule{2-16} \multicolumn{1}{|c|}{} & \multicolumn{1}{c|}{\multirow{6}[4]{*}{UMAP}} & \multicolumn{1}{c|}{\multirow{3}[2]{*}{DE}} & \multicolumn{1}{c|}{CC- Stg 2} & \multicolumn{1}{c}{\multirow{3}[2]{*}{K-means}} & \multicolumn{1}{c}{\multirow{3}[2]{*}{k~=~30}} & \multicolumn{1}{c}{\multirow{3}[2]{*}{CHI - 13293,156}} & \multicolumn{1}{c|}{\multirow{3}[2]{*}{30}} & \multicolumn{1}{c}{hgpa} & \multicolumn{1}{c}{HGPA} & \multicolumn{1}{c|}{163} & \multicolumn{1}{c}{0} & \multicolumn{1}{c}{0.37} & \multicolumn{1}{c}{6} & \multicolumn{1}{c}{2.20} & \multicolumn{1}{c|}{5} \\
 \multicolumn{1}{|c|}{} & \multicolumn{1}{c|}{} & \multicolumn{1}{c|}{} & \multicolumn{1}{c|}{HM - Stg 2} &            &            &            & \multicolumn{1}{c|}{} & \multicolumn{1}{c}{Euc\_15\_20} & \multicolumn{1}{c}{HDBSCAN} & \multicolumn{1}{c|}{63} & \multicolumn{1}{c}{0} & \multicolumn{1}{c}{0.37} & \multicolumn{1}{c}{6} & \multicolumn{1}{c}{2.24} & \multicolumn{1}{c|}{6} \\
 \multicolumn{1}{|c|}{} & \multicolumn{1}{c|}{} & \multicolumn{1}{c|}{} & \multicolumn{1}{c|}{ANMI - Stg 2} &            &            &            & \multicolumn{1}{c|}{} & \multicolumn{1}{c}{L1\_Avg\_10} & \multicolumn{1}{c}{Agg} & \multicolumn{1}{c|}{10} & \multicolumn{1}{c}{0} & \multicolumn{1}{c}{0.20} & \multicolumn{1}{c}{6} & \multicolumn{1}{c}{1.23} & \multicolumn{1}{c|}{3} \\
\cmidrule{3-16} \multicolumn{1}{|c|}{} & \multicolumn{1}{c|}{} & \multicolumn{1}{c|}{\multirow{3}[2]{*}{AE}} & \multicolumn{1}{c|}{CC- Stg 1} & \multicolumn{1}{c}{\multirow{3}[2]{*}{K-means}} & \multicolumn{1}{c}{\multirow{3}[2]{*}{k~=~30}} & \multicolumn{1}{c}{\multirow{3}[2]{*}{CHI - 13293,156}} & \multicolumn{1}{c|}{\multirow{3}[2]{*}{30}} & \multicolumn{1}{c}{hgpa} & \multicolumn{1}{c}{HGPA} & \multicolumn{1}{c|}{250} & \multicolumn{1}{c}{0} & \multicolumn{1}{c}{0.34} & \multicolumn{1}{c}{6} & \multicolumn{1}{c}{2.04} & \multicolumn{1}{c|}{4} \\
 \multicolumn{1}{|c|}{} & \multicolumn{1}{c|}{} & \multicolumn{1}{c|}{} & \multicolumn{1}{c|}{HM - Stg 1} &            &            &            & \multicolumn{1}{c|}{} & \multicolumn{1}{c}{Birch\_30} & \multicolumn{1}{c}{Birch} & \multicolumn{1}{c|}{30} & \multicolumn{1}{c}{0} & \multicolumn{1}{c}{0.13} & \multicolumn{1}{c}{6} & \multicolumn{1}{c}{0.77} & \multicolumn{1}{c|}{\cellcolor[rgb]{ .851,  .851,  .851}\textbf{1}} \\
 \multicolumn{1}{|c|}{} & \multicolumn{1}{c|}{} & \multicolumn{1}{c|}{} & \multicolumn{1}{c|}{ANMI - Stg 1} &            &            &            & \multicolumn{1}{c|}{} & \multicolumn{1}{c}{Euc\_avg\_24} & \multicolumn{1}{c}{Agg} & \multicolumn{1}{c|}{24} & \multicolumn{1}{c}{0} & \multicolumn{1}{c}{0.14} & \multicolumn{1}{c}{6} & \multicolumn{1}{c}{0.83} & \multicolumn{1}{c|}{2} \\
\cmidrule{2-16} \multicolumn{1}{|c|}{} & \multicolumn{1}{c|}{\multirow{6}[4]{*}{\textbf{N2d\_2d}}} & \multicolumn{1}{c|}{\multirow{3}[2]{*}{DE}} & \multicolumn{1}{c|}{CC- Stg 2} & \multicolumn{1}{c}{\multirow{3}[2]{*}{K-means}} & \multicolumn{1}{c}{\multirow{3}[2]{*}{k~=~30}} & \multicolumn{1}{c}{\multirow{3}[2]{*}{CHI - 9998,848}} & \multicolumn{1}{c|}{\multirow{3}[2]{*}{30}} & \multicolumn{1}{c}{hgpa} & \multicolumn{1}{c}{HGPA} & \multicolumn{1}{c|}{163} & \multicolumn{1}{c}{0} & \multicolumn{1}{c}{0.27} & \multicolumn{1}{c}{6} & \multicolumn{1}{c}{1.63} & \multicolumn{1}{c|}{5} \\
 \multicolumn{1}{|c|}{} & \multicolumn{1}{c|}{} & \multicolumn{1}{c|}{} & \multicolumn{1}{c|}{HM - Stg 2} &            &            &            & \multicolumn{1}{c|}{} & \multicolumn{1}{c}{L2\_15\_10} & \multicolumn{1}{c}{HDBSCAN} & \multicolumn{1}{c|}{112} & \multicolumn{1}{c}{0} & \multicolumn{1}{c}{0.27} & \multicolumn{1}{c}{6} & \multicolumn{1}{c}{1.62} & \multicolumn{1}{c|}{4} \\
 \multicolumn{1}{|c|}{} & \multicolumn{1}{c|}{} & \multicolumn{1}{c|}{} & \multicolumn{1}{c|}{ANMI - Stg 2} &            &            &            & \multicolumn{1}{c|}{} & \multicolumn{1}{c}{L2\_20\_20} & \multicolumn{1}{c}{HDBSCAN} & \multicolumn{1}{c|}{80} & \multicolumn{1}{c}{0} & \multicolumn{1}{c}{0.28} & \multicolumn{1}{c}{6} & \multicolumn{1}{c}{1.67} & \multicolumn{1}{c|}{6} \\
\cmidrule{3-16} \multicolumn{1}{|c|}{} & \multicolumn{1}{c|}{} & \multicolumn{1}{c|}{\multirow{3}[2]{*}{\textbf{AE}}} & \multicolumn{1}{c|}{CC- Stg 1} & \multicolumn{1}{c}{\multirow{3}[2]{*}{K-means}} & \multicolumn{1}{c}{\multirow{3}[2]{*}{k~=~30}} & \multicolumn{1}{c}{\multirow{3}[2]{*}{CHI - 9998,848}} & \multicolumn{1}{c|}{\multirow{3}[2]{*}{30}} & \multicolumn{1}{c}{hgpa} & \multicolumn{1}{c}{HGPA} & \multicolumn{1}{c|}{276} & \multicolumn{1}{c}{0} & \multicolumn{1}{c}{0.28} & \multicolumn{1}{c}{\cellcolor[rgb]{ .851,  .851,  .851}\textcolor[rgb]{ 1,  0,  0}{\textbf{2}}} & \multicolumn{1}{c}{0.55} & \multicolumn{1}{c|}{2} \\
 \multicolumn{1}{|c|}{} & \multicolumn{1}{c|}{} & \multicolumn{1}{c|}{} & \multicolumn{1}{c|}{\cellcolor[rgb]{ .851,  .851,  .851}\textbf{HM - Stg 1}} &            &            &            & \multicolumn{1}{c|}{} & \multicolumn{1}{c}{\textbf{Agg\_ward\_30}} & \multicolumn{1}{c}{\textbf{Agg}} & \multicolumn{1}{c|}{\textbf{30}} & \multicolumn{1}{c}{\cellcolor[rgb]{ .851,  .851,  .851}\textbf{1}} & \multicolumn{1}{c}{\textbf{0.00}} & \multicolumn{1}{c}{\cellcolor[rgb]{ .851,  .851,  .851}\textbf{2}} & \multicolumn{1}{c}{\textbf{0.00}} & \multicolumn{1}{c|}{\cellcolor[rgb]{ .851,  .851,  .851}\textbf{1}} \\
 \multicolumn{1}{|c|}{} & \multicolumn{1}{c|}{} & \multicolumn{1}{c|}{} & \multicolumn{1}{c|}{ANMI - Stg 1} &            &            &            & \multicolumn{1}{c|}{} & \multicolumn{1}{c}{Agg\_euc\_avr\_28} & \multicolumn{1}{c}{Agg} & \multicolumn{1}{c|}{28} & \multicolumn{1}{c}{0} & \multicolumn{1}{c}{0.15} & \multicolumn{1}{c}{6} & \multicolumn{1}{c}{0.88} & \multicolumn{1}{c|}{3} \\
 \midrule
 \midrule
 \multicolumn{16}{p{95.925em}}{Abbreviations used in the table: $Strategy w/Alg$ \textbf{Decreased Ensemble}: DA, \textbf{All Ensemble}: AE, \textbf{Consensus Clustering}: CC, \textbf{Hyperparameter Match}: HM, \textbf{Averaged Normalized Mutual Information}: ANMI, $Strategy$: Stg, $Hyper-parameters$: \textbf{Complete}: comp, \textbf{Average}: avg, \textbf{Euclidean}: Euc, $Class$: \textbf{Agglomerative}: Agg} \\
 \end{tabular}%
\label{tab:filteringapproach}%
\end{sidewaystable}

\begin{figure*}
    \centering
    \vspace{-0.5pt}
    \includegraphics[width=15cm, height=19cm]{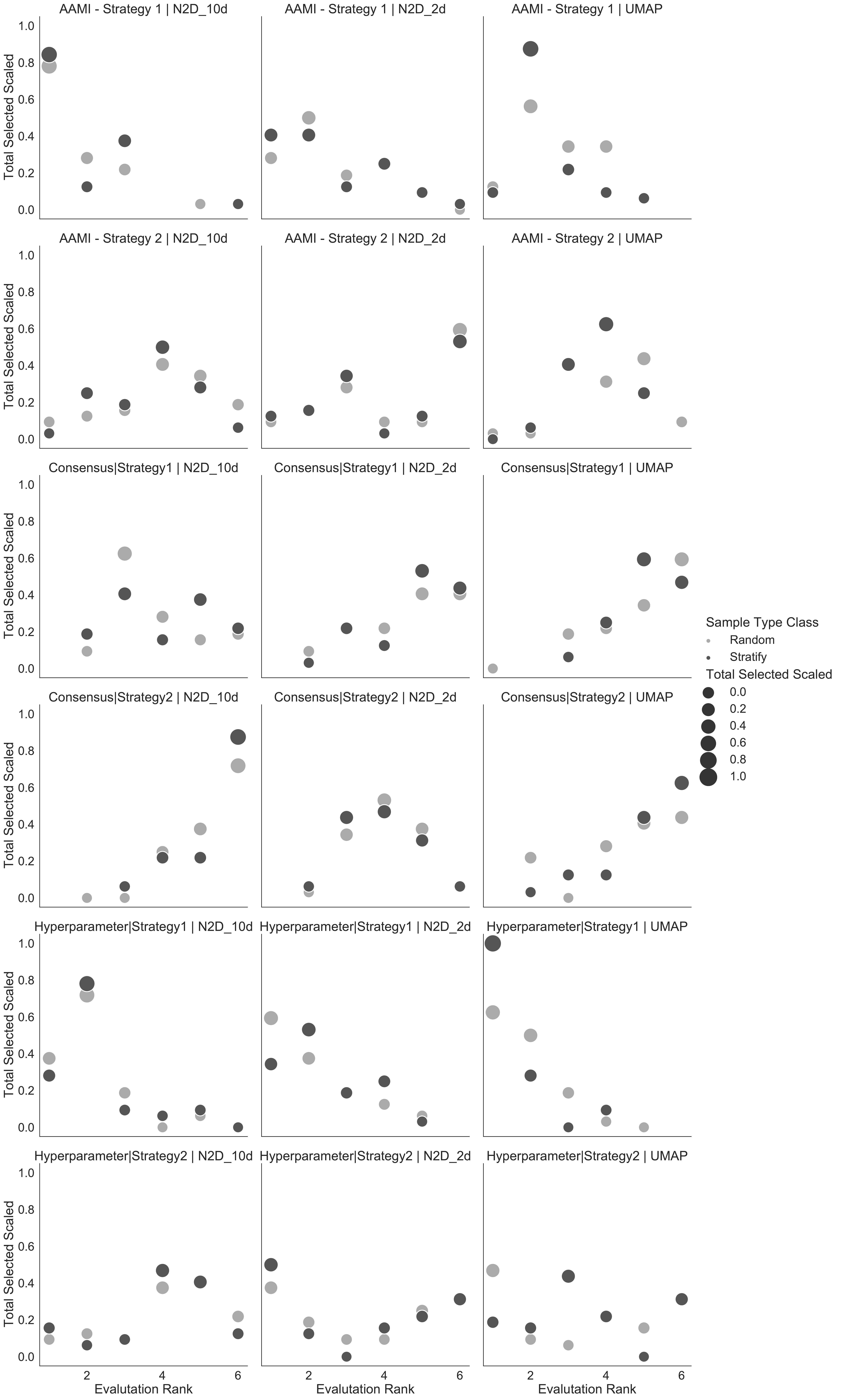}
    \caption{Closeness of the metric scores obtained with the applied strategies to the optimal metric scores selected individually for the sampled dataset and ranked according to the filtering approach. Each embedding is represented in columns as N2d-10d, N2d-2d, and UMAP. While the columns show the embeddings, each strategy applied is represented in the rows. Stratify and random sample type breakdowns are discriminated in black and grey colours, respectively. Each chart includes rankings from 1 to 6 on the X-axis, while the Y-axis shows the total selections of these rankings during the year. The choice closest to the optimal metric value shows the approach and strategy with the highest total selection value in the $1^{st}$ rank in this chart.}
    \label{fig:approach_success}
\end{figure*}

As illustrated in Fig~\ref{fig:approach_success}, even though there are ranked one results and one-to-one matching algorithms with optimal algorithms of N2d-10d and N2d-2d embeddings, the closest approach to the optimal algorithm is Hyperparameter Match Strategy 1 with UMAP embedding, which has the highest number of selections in rank 1. Based on these results, we conclude that Hyperparameter Match Strategy 1 with the UMAP result is the closest approach to the optimal values. Additionally, we conclude that the CHI metric produces the best quality clusters compared to other metrics. When we compare sample types, we can say that stratifying and random sampling mostly show similar behaviours, which may be due to the solid semantic correlation of the datasets within months.

\subsubsection{Cluster Stability} \label{ClusterStability}
The stability of the clusters is defined as the reoccurrence of the resulting clusters in different periods, even when the data points change. Some previous studies applied this procedure by having frequent samples without replacements and averaging the similarities with the Jaccard coefficient~\citep{clustertend_DanielGrech}. Another study investigated this approach to compare similarity between monthly samples of a digital library~\citep{TesselLaura} which concluded that there is a trade-off between silhouette width and the stability scores of the clusters. Our study validates the stability of resulting clusters in separated months by computing the average AMI scores to indicate clustering quality, which also means that these clusters are the general reading behaviours of the users whose referral channels are~Twitter.

\textbf{Cluster Distributions:} To determine the stability, for each cluster, $C_i$, the distribution of the categorized and binned feature values, $f_k$, of $n$ dimensional feature set, $F_i^j~=~\{F_i^1, F_i^2,...F_i^n\}$, is obtained in each sample as follows: 
\begin{equation}
	F_i^j~=~\Bigg(\dfrac{f_1}{C_{i}},\dfrac{f_2}{C_{i}},\dots,\dfrac{f_k}{C_{i}}\Bigg), \quad {C_{i}~=~}\sum_{k}{f_{k}},
	\label{profiledistr}
\end{equation}	
where $F_i^j$ represents the $j^{th}$ feature of the $i^{th}$ cluster, and the size of the cluster $C_i$ is defined by the total size of features, $f_k$.

\begin{figure*}
    \centering
    \vspace{-0.5pt}
    \includegraphics[width=17cm, height=9cm]{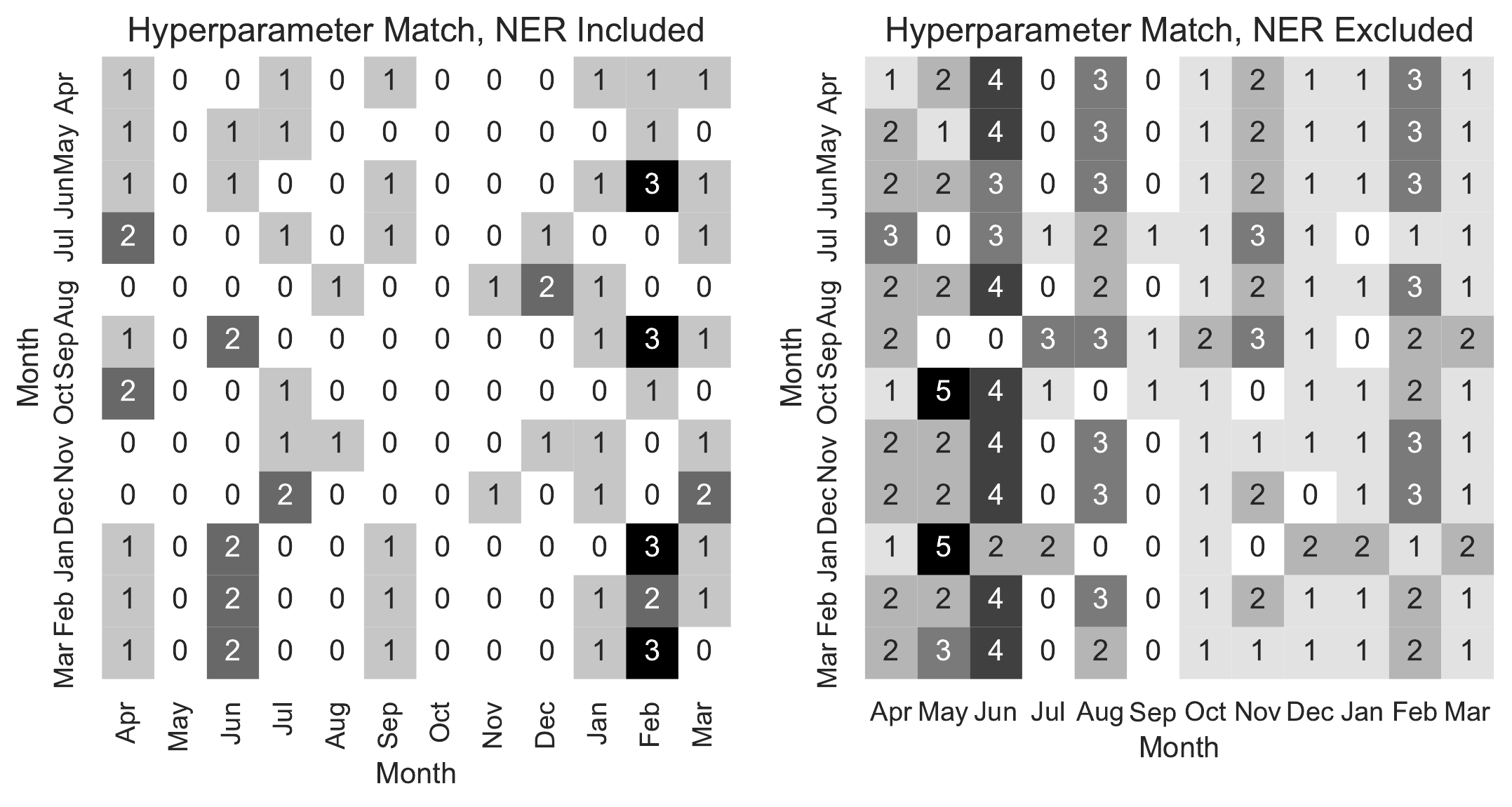}
    \caption{Comparison of all feature distributions over all clusters with AMI. The left figure includes the NER results, which are excluded from the right figure.}
    \label{fig:cluster_stability_over_months}
\end{figure*}

Fig~\ref{fig:cluster_stability_over_months} shows, per cluster, that have more than 0.5 AMI values according to \ref{profiledistr} when compared with the rest of the clusters in all months that are labelled  as high stability. Since NER attributes are mostly related to the given text of articles, they are compared by including and excluding these features. The results show that the highest stability counts can be obtained without NER attributes with an average stability of 10 out of 12 months. For example, in the Hyperparameter Match strategy, while 1 cluster obtained from April has matching clusters in June with NER included, the results increase to 4 when NER results are excluded.

Table~\ref{tab:monthly_reoccurance_clusters} represents the characteristics of the clusters related to the specific month, which reoccurred at the highest rate in other months. The observations from these clusters can be summarized as follows:
\begin{itemize}
\item In particular, the recently published articles that bounced from the first news item referred to in $City1$ are stable for all clusters.
\item The clusters with the highest volume did not achieve the similarity threshold with any clusters; instead, when we observe over months, the stability of the results is high, with a minimum rank of 4 and a maximum of 20 among all clusters.
\item The entertainment category is primarily read in the afternoon with mobile devices.
\item While weekday readers continue reading steadily over the months, the weekend readerships do not have high stability rates.
\item The clusters that have columnists have higher ranked articles than other categories.
\item While mobile users are often the greatest among device selection, PC users mostly read current affairs with political issues.
\end{itemize}

 \begin{table*}[align, pos=h!, hbt!] 
   \caption{Clusters with the highest reoccurrence count for a specific month}
 \begin{adjustbox}{width=\textwidth}
     
 \begin{tabular}{p{9.085em}llllllllllll}
 \multicolumn{13}{c}{} \\
 \midrule
 \textbf{Month} & \multicolumn{1}{p{3.25em}}{\textbf{Apr}} & \multicolumn{1}{p{3.915em}}{\textbf{May}} & \multicolumn{1}{p{3.25em}}{\textbf{Jun}} & \multicolumn{1}{p{3.25em}}{\textbf{Jul}} & \multicolumn{1}{p{3.585em}}{\textbf{Aug}} & \multicolumn{1}{p{3.665em}}{\textbf{Sep}} & \multicolumn{1}{p{3.75em}}{\textbf{Oct}} & \multicolumn{1}{p{3em}}{\textbf{Nov}} & \multicolumn{1}{p{3.75em}}{\textbf{Dec}} & \multicolumn{1}{p{3.25em}}{\textbf{Jan}} & \multicolumn{1}{p{3.75em}}{\textbf{Feb}} & \multicolumn{1}{p{3.25em}}{\textbf{Mar}} \\
 \midrule
 \midrule
 \textbf{Cluster Size(\%) - Rank} & \multicolumn{1}{p{3.25em}}{4,36 - 6} & \multicolumn{1}{p{3.915em}}{2,84 - 20} & \multicolumn{1}{p{3.25em}}{5,87 - 4} & \multicolumn{1}{p{3.25em}}{2,4 - 19} & \multicolumn{1}{p{3.585em}}{3,98 - 12} & \multicolumn{1}{p{3.665em}}{2,19 - 20} & \multicolumn{1}{p{3.75em}}{4,38 - 10} & \multicolumn{1}{p{3em}}{4,4 - 9} & \multicolumn{1}{p{3.75em}}{4,35 - 11} & \multicolumn{1}{p{3.25em}}{5,69 - 4} & \multicolumn{1}{p{3.75em}}{2,42 - 24} & \multicolumn{1}{p{3.25em}}{4,33 - 8} \\
 \midrule
 \textbf{\#of Reoccurances - Unique Months } & \multicolumn{1}{p{3.25em}}{19 - 10} & \multicolumn{1}{p{3.915em}}{19 - 10} & \multicolumn{1}{p{3.25em}}{19 - 10} & \multicolumn{1}{p{3.25em}}{17 - 10} & \multicolumn{1}{p{3.585em}}{19 - 10} & \multicolumn{1}{p{3.665em}}{19 - 9} & \multicolumn{1}{p{3.75em}}{18 - 10} & \multicolumn{1}{p{3em}}{19 - 10} & \multicolumn{1}{p{3.75em}}{19 - 9} & \multicolumn{1}{p{3.25em}}{18 - 9} & \multicolumn{1}{p{3.75em}}{19 - 10} & \multicolumn{1}{p{3.25em}}{18 - 10} \\
 \midrule
 \textbf{Algorithm - \#of Clusters} & \multicolumn{1}{p{3.25em}}{B - 30} & \multicolumn{1}{p{3.915em}}{B - 26} & \multicolumn{1}{p{3.25em}}{A - 30} & \multicolumn{1}{p{3.25em}}{B - 30} & \multicolumn{1}{p{3.585em}}{B - 30} & \multicolumn{1}{p{3.665em}}{A - 30} & \multicolumn{1}{p{3.75em}}{A - 30} & \multicolumn{1}{p{3em}}{A - 30} & \multicolumn{1}{p{3.75em}}{B - 20} & \multicolumn{1}{p{3.25em}}{A - 30} & \multicolumn{1}{p{3.75em}}{A - 30} & \multicolumn{1}{p{3.25em}}{A - 30} \\
 \midrule
 \textbf{Bounce(\%)} & \multicolumn{1}{c}{86.7} & \multicolumn{1}{c}{97.18} & \multicolumn{1}{c}{93.9} & \multicolumn{1}{c}{94.17} & \multicolumn{1}{c}{95.98} & \multicolumn{1}{c}{92.66} & \multicolumn{1}{c}{89.55} & \multicolumn{1}{c}{80} & \multicolumn{1}{c}{73.73} & \multicolumn{1}{c}{55.48} & \multicolumn{1}{c}{90.91} & \multicolumn{1}{c}{96.77} \\
 \midrule
 \textbf{City1(\%)} & \multicolumn{1}{c}{44.95} & \multicolumn{1}{c}{44.37} & \multicolumn{1}{c}{41.69} & \multicolumn{1}{c}{47.5} & \multicolumn{1}{c}{46.73} & \multicolumn{1}{c}{43.12} & \multicolumn{1}{c}{48.18} & \multicolumn{1}{c}{40.91} & \multicolumn{1}{c}{47.93} & \multicolumn{1}{c}{48.76} & \multicolumn{1}{c}{41.32} & \multicolumn{1}{c}{45.62} \\
 \midrule
 \textbf{Recently Published (\%)} & \multicolumn{1}{c}{97.25} & \multicolumn{1}{c}{93.66} & \multicolumn{1}{c}{99.66} & \multicolumn{1}{c}{98.33} & \multicolumn{1}{c}{95.98} & \multicolumn{1}{c}{95.41} & \multicolumn{1}{c}{86.82} & \multicolumn{1}{c}{98.64} & \multicolumn{1}{c}{98.62} & \multicolumn{1}{c}{98.23} & \multicolumn{1}{c}{28.1} & \multicolumn{1}{c}{60.37} \\
 \midrule
 \textbf{Ranking(\%)} & \multicolumn{1}{p{3.25em}}{LR - 78,9} & \multicolumn{1}{p{3.915em}}{LR - 40,14} & \multicolumn{1}{p{3.25em}}{MR - 68,47} & \multicolumn{1}{p{3.25em}}{MR - 31,67} & \multicolumn{1}{p{3.585em}}{LR - 43,72} & \multicolumn{1}{p{3.665em}}{LR - 41,28} & \multicolumn{1}{p{3.75em}}{LR - 45,45} & \multicolumn{1}{p{3em}}{LR - 41,82} & \multicolumn{1}{p{3.75em}}{LR - 36,41} & \multicolumn{1}{p{3.25em}}{LR - 33,22} & \multicolumn{1}{p{3.75em}}{LR - 97,52} & \multicolumn{1}{p{3.25em}}{LR - 97,24} \\
 \midrule
 \textbf{Category(\%)} & \multicolumn{1}{p{3.25em}}{CA - 63,3} & \multicolumn{1}{p{3.915em}}{E - 71,13} & \multicolumn{1}{p{3.25em}}{C - 56,95} & \multicolumn{1}{p{3.25em}}{C - 35} & \multicolumn{1}{p{3.585em}}{Et - 91,46} & \multicolumn{1}{p{3.665em}}{CA - 59,63} & \multicolumn{1}{p{3.75em}}{CA - 29,09} & \multicolumn{1}{p{3em}}{Et - 32,27} & \multicolumn{1}{p{3.75em}}{CA - 89,86} & \multicolumn{1}{p{3.25em}}{CA - 79,86} & \multicolumn{1}{p{3.75em}}{Et - 33,88} & \multicolumn{1}{p{3.25em}}{T - 38,71} \\
 \midrule
 \textbf{Device(\%)} & \multicolumn{1}{p{3.25em}}{M - 59,63} & \multicolumn{1}{p{3.915em}}{M - 97,89} & \multicolumn{1}{p{3.25em}}{M - 70,85} & \multicolumn{1}{p{3.25em}}{M - 75,83} & \multicolumn{1}{p{3.585em}}{M - 85,43} & \multicolumn{1}{p{3.665em}}{M - 82,57} & \multicolumn{1}{p{3.75em}}{M - 57,73} & \multicolumn{1}{p{3em}}{PC - 76,82} & \multicolumn{1}{p{3.75em}}{PC - 94,01} & \multicolumn{1}{p{3.25em}}{PC - 78,8} & \multicolumn{1}{p{3.75em}}{M - 87,6} & \multicolumn{1}{p{3.25em}}{M - 82,49} \\
 \midrule
 \textbf{Hour(\%)} & \multicolumn{1}{p{3.25em}}{AN - 9,63} & \multicolumn{1}{p{3.915em}}{AN - 11,27} & \multicolumn{1}{p{3.25em}}{MR - 13,22} & \multicolumn{1}{p{3.25em}}{MR - 21,67} & \multicolumn{1}{p{3.585em}}{AN - 13,07} & \multicolumn{1}{p{3.665em}}{AN - 17,43} & \multicolumn{1}{p{3.75em}}{MR - 14,09} & \multicolumn{1}{p{3em}}{AN - 14,09} & \multicolumn{1}{p{3.75em}}{MR - 12,9} & \multicolumn{1}{p{3.25em}}{AN - 14,84} & \multicolumn{1}{p{3.75em}}{AN - 10,74} & \multicolumn{1}{p{3.25em}}{MR - 15,67} \\
 \midrule
 \textbf{NER(\%)} & \multicolumn{1}{p{3.25em}}{IP - 17,4} & \multicolumn{1}{p{3.915em}}{UNK - 46,71} & \multicolumn{1}{p{3.25em}}{P - 26,44} & \multicolumn{1}{p{3.25em}}{P - 13,33} & \multicolumn{1}{p{3.585em}}{EtS - 63,82} & \multicolumn{1}{p{3.665em}}{UNK - 42,81} & \multicolumn{1}{p{3.75em}}{None 21,6} & \multicolumn{1}{p{3em}}{UNK - 22,12} & \multicolumn{1}{p{3.75em}}{UNK - 10,6} & \multicolumn{1}{p{3.25em}}{ P - 13,07} & \multicolumn{1}{p{3.75em}}{IP - 18,18} & \multicolumn{1}{p{3.25em}}{UNK - 59,44} \\
 \midrule
 \textbf{Weekday(\%)} & \multicolumn{1}{c}{39.45} & \multicolumn{1}{c}{42.25} & \multicolumn{1}{c}{26.1} & \multicolumn{1}{c}{49.17} & \multicolumn{1}{c}{26.63} & \multicolumn{1}{c}{56.88} & \multicolumn{1}{c}{25.45} & \multicolumn{1}{c}{34.09} & \multicolumn{1}{c}{23.5} & \multicolumn{1}{c}{80.21} & \multicolumn{1}{c}{40.5} & \multicolumn{1}{c}{39.63} \\
 \midrule
 \textbf{Weeknumber(\%)} & \multicolumn{1}{p{3.25em}}{$1^{st}$ - 78,44} & \multicolumn{1}{p{3.915em}}{$3^{rd}$ - 78,17} & \multicolumn{1}{p{3.25em}}{$2^{nd}$ - 98,64} & \multicolumn{1}{p{3.25em}}{$3^{rd}$ - 58,33} & \multicolumn{1}{p{3.585em}}{$4^{th}$ - 100,0} & \multicolumn{1}{p{3.665em}}{$4^{th}$ - 96,33} & \multicolumn{1}{p{3.75em}}{$2^{nd}$ - 46,82} & \multicolumn{1}{p{3em}}{$2^{nd}$ - 33,18} & \multicolumn{1}{p{3.75em}}{$4^{th}$ - 96,31} & \multicolumn{1}{p{3.25em}}{$4^{th}$ - 50,18} & \multicolumn{1}{p{3.75em}}{$3^{rd}$ - 70,25} & \multicolumn{1}{p{3.25em}}{$2^{nd}$ - 60,37} \\
 \midrule
 \midrule
 \multicolumn{13}{p{65.75em}}{Abbreviations used in the table: $Algorithms$ \textbf{Agglomerative}: A; \textbf{Birch}: B; $Rank$  \textbf{Low Rank}: LR; \textbf{Mid Rank}: MR; $Category$  \textbf{Current Affairs}: CA; \textbf{Economy}: E; \textbf{Entertainment}: Et; \textbf{Travel}: T; \textbf{Columnist}: C; $Device$  \textbf{Mobile}: M; $Hour$   \textbf{Afternoon}: AN; \textbf{Morning}: MR;  $NER$  \textbf{City1 - Politics}: IP; \textbf{Unknown}: UNK; \textbf{Politics} : P; \textbf{Entertainment show}: EtS} \\
 \end{tabular}%

   \label{tab:monthly_reoccurance_clusters}%
 \end{adjustbox}
\end{table*}

\section{Conclusions and Future Work} \label{conclusion}
Based on detailed clickstream data, this study reveals the reading patterns of anonymous readers whose referral channel is~Twitter. Reading characteristics are first analysed by the $lead category$ distribution over months by location and device breakdowns, demonstrating that the underlying distribution diverges over time. An ensemble clustering analysis is utilized by comparing different embedding approaches and strategies on methodologically selected samples to identify stable clusters independent of time. Since the methodologies are applied based on the data from one news outlet, H\"{u}rriyet, the results may not be generalizable; thus, we conduct the stability analysis over months, which yields profoundly stable patterns. The promising characteristics of these findings show that some columnists have a substantial and continuous impact on delivering readers to a website. Mostly low ranked news is found to be directed, but in regard to the columnists and rarely directed locations, the behaviours change mainly to the high ranked contents. In general, morning readers are the majority, but entertainment news is mostly read in the afternoon. The latest generated news is preferred primarily by the readers, with which we can deduce that the readers are most interested in trending news.

Our empirical analyses with a comprehensive perspective of the clustering methodologies could find stable reading habits, leading to a number of further investigations. Our first plan is to reduce the complexity and increase the robustness of the experimental study on the decreased ensemble sets that we conducted, which can be improved by an iterative approach to find the optimal thresholds. We plan to incorporate sequential article data on news consumption, which is absent in our methodologies and could be insufficient to find the impacts of consecutive reading behaviours. Since our study focus does not fully reveal the underlying mechanism of these referral effects, future work based on social media with a holistic view may broadly expose these channels’ reading characteristics. To enrich the semantics of future work, we plan to study transformer-based networks to explore the success of clickstream data embeddings on news outlet readership.

\clearpage

\appendix
\renewcommand{\thesection}{\Alph{section}.\arabic{section}}

\section{Clustering Problem} \label{Appendix}
The main goal of the clustering problem is to include similar data points in the same cluster and include different data points in separated clusters~\citep{cls_Jain}. According to this definition, we can define the clustering problem as follows in order to get the optimal result: Given set of data with $n$ data points in $d$ dimensions $D~=~\{x_1,x_2,...x_n\} \subseteq\ R^{d}$ we are in search of $k$ number of clusters $C~=~\{c_1,c_2,...c_k\}$ whose intersection sets are empty and represents the partitions of primary data set. As a result, each data point is expressed with a label. When comparing clusters, one of the essential consideration is that labels obtained from different clustering algorithms can give the same similarity result for data points included in the same cluster, even if they come in different permutations. Because the purpose here is to compare whether the same data points are in the same cluster or different clusters. When we observe the labels obtained from two different clustering algorithms, such as $K$ and $K^{'}$, we aim to return the similarity ratio between [1,1,2,3,3] and [3,3,1,2,2] to be the maximum AMI value

\section{Internal Validation Metrics}
\setlength{\parskip}{\baselineskip}%
\setlength{\parindent}{0pt}%

\textbf{Silhouette coefficient (SI):} This metric gives results between [-1,1], that is, the similarity of the data points within the cluster, $a(i)$, compared to other clusters, $b(i)$, where d(i,j) is the distance metric. Higher values indicate appropriate clusters.

\begin{equation}
    {a(i)~=~} \dfrac{1}{|C_i|-1}\sum_{j~\in C, i\neq j} {d(i,j)}
	\notag
\end{equation}

\begin{equation}
    {b(i)~=~} min_{k\neq j} \dfrac{1}{|C_k|} \sum_{j~\in C_k} {d(i,j)}
	\notag
\end{equation}

\begin{equation}
	{SI~=~}\dfrac{1}{|D|}\sum_{i~\in D} \dfrac {b(i) - a(i)}{max\{a(i),b(i)\}}
\end{equation}

\textbf{Dunn index-type (DI):} The value is a given ratio of inter-cluster distances over intra-cluster diameters, where higher values indicate better clusters. Out of several approaches to define the diameter of the cluster, in this study, we used the mean of pairwise distances in the cluster. Assuming $x$ and $y$ are the data points assigned in the same cluster from the given dataset $D$ in $n$ dimensions and there are $k$ clusters, pairwise distances mean calculated as follows:

\begin{equation}
	{\mu_ =}~\dfrac{\sum_{x~\in C_i} {x}}{|C_i|},   {D_m =}~\dfrac{\sum_{x~\in~C_i}d(x,\mu)}{|C_i|}
	\notag
\end{equation}

\begin{equation}
    {DI_k~=~}\dfrac{min_{1\leq i\leq j \leq k} \delta({C_i},{C_j})}{max_{1\leq l\leq k} D_k}
\end{equation}

\textbf{Calinski Harabasz index (CHI):} This index aims to find the ratio between cluster variance over within cluster variance, which is in the range of $[0,\infty)$. Higher values indicate good clustering.

\begin{equation}
	{CHI~=~}\dfrac {\dfrac{1}{k-1}\sum_{i~=~1}^{k}{||\mu_i - \mu||^2}} {\dfrac{1}{|D|-1} \sum_{i~=~1}^{k}\sum_{{x}~\in {C_i}} {||{x} - \mu||^2}}
\end{equation}

\textbf{Davies Bouldin scores (DB):} This score gives how similar each cluster $(C_i)$ is to its most similar cluster $(C_j)$ on average. It measures intra- and inter-cluster distance ratios. The lower values indicate appropriate clustering where it ranges between $[0,\infty)$. When the average distance between each data point and the centre of the cluster is defined as ${S_i}$ and the inter-cluster centroid distance is defined as ${N_{i,j}}$, the measurement of the goodness of the cluster, ${R_{i,j}}$, and the Davies-Bouldin index is defined as follows:

\begin{equation}
	{R_{i,j}~=~}\dfrac{S_i + S_j}{N_{i,j}}
	\notag
\end{equation}

\begin{equation}
	{DB~=~}\dfrac{1}{k}\sum_{i=1}^{k} {max_{i\neq j}{R_{i,j}}}
\end{equation}

\textbf{Scatter and Density between clusters (S-Dbw) Validity Index:} This metric consists of two objectives, as shown below. The first part ${Scat(\#c)}$ is related to the average scattering within the clusters that is obtained by partitioning the dataset. Therefore, minimum values indicate better results in terms of compactness and separation. The second part $Dens\_bw(\#c)$ considers intra-cluster densities by comparing the clusters’ centre points; thus, minimum values also indicate better separation. 

\begin{equation}
	{Scat(\#c)~=~}\dfrac {\dfrac{1}{\#c}\sum_{i~=~1}{\left\|\sigma(C_i)\right\|}}{\left\|\sigma(D)\right\|}
	\notag
\end{equation}

\begin{multline*}
	{Dens\_bw(\#c)~=~}\dfrac {1}{\#c(\#c-1)} \sum_{i~=~1}\\\Big[\sum_{j,j\neq1}   \dfrac{\sum_{x~\in C_i \cup C_j} {f(x,u_{i,j})} } {max\{\sum_x~\in C_i f(x, c_i), \sum_x~\in C_j f(x, c_j)\}}\Big]
	\notag
\end{multline*}

\begin{equation}
	{S-Dbw~=~}{Scat(\#c)} + {Dens\_bw(\#c))}
\end{equation}

($\#c$:Number of total clusters, $C_{i}$:i-th cluster,~$\sigma(C_{i})$:variance of $C_{i}$,~$D$:Dataset,~ $f(x,u_{i,j})$: the midpoint of the clusters' centers as a defined line segment, ~$\left\|X\right\|~=~(X^{T}X)^{1/2}$)     

There are other versions on S-Dbw which we compare during the study. The problem with S-Dbw reflected by other studies. The first drawback is, since Dens\_bw(\#c) in S-Dbw measures neighborhood within the standard deviation, it could not cover the inter-cluster similarity especially for the non-circular clusters. Secondly, Scat(\#c) part for intra-cluster similarity does not take into account tuples which might effect when there is similar variance among two cluster with difference tuple sizes. 

The approach introduced by Kim et. al.~\citep{Sdbw_Kim}, for the second part of the formulation, uses cluster based inter-cluster similarity by considering confidence interval for each dimension which provides to be defined the size and shape of the densities of clusters, on the other hand for intra-cluster similarity uses weighted average variance within clusters.     

Tong et. al~\citep{Sdbw_Tong} note some drawbacks introduced by~\citep{Sdbw_Kim}. Since~\citep{Sdbw_Kim} does not consider the discrepancy among the clusters, selecting the middle point of the margin notoriously affects the centre choice. Scat(\#c) points out that when cluster numbers are increasing, the result of this part will be monotonically increasing as well. To overcome these, for the Dens\_bw(\#c) part, this approach selects the margin region instead of the middle point of the line segment among clusters to improve intercluster similarity. For intracluster similarity. Within-cluster similarity prevents monotonically increasing values by keeping the constant, although the number of clusters increases.

\section{External Validation Metrics}
\setlength{\parskip}{\baselineskip}%
\setlength{\parindent}{0pt}%
\textbf{Normalized Mutual Information (NMI):} In information theory, Mutual Information (MI) is defined as the amount of information obtained from one discrete-valued random variable $X$, by observing the other random variable $Y$. Assuming the joint probability distribution of two random variables is $\rho_{X,Y}$ and the expected value of this distribution is $\mathbf{E_\rho}$, then MI is defined as:

\begin{equation}
	{I(X,Y)~=~} \mathbf{E_{\rho_{X,Y}}} \Bigg[\log \dfrac{\rho_{X,Y}}{\rho_{X},\rho_{Y}}\Bigg]
	\notag
\end{equation}

Since $I(X,Y)$ is not bounded above, normalizing it provides the results in the range of $[0,1]$, assuming ${H(X)}$ and ${H(Y)}$ are the marginal entropies of random variables;

\begin{equation}
	{H(X)~=~} -\mathbf{E\rho_{X}} [\log \rho_{X}] , {I(X,Y)}\leq min{(H(X),H(Y))}
	\notag
\end{equation}

\begin{equation}
	{NMI(X,Y)~=~} \dfrac{I(X,Y)}{\sqrt{H(X)H(Y)}}
\end{equation}

As defined in~\citep{ensembles_Ghosh}, NMI can be interpreted as the label similarities of the clustering results, even though they might come from different permutations. In this case, the assumed X and Y discrete random variables can be interpreted as ${C}$, as clustered data label assignments, and ${C^{*}}$  as ground-truth labels.

\textbf{Adjusted Mutual Information (AMI):} The adjustment of the result from MI is needed, since MI values are higher when there are a larger number of clusters not necessarily having more information shared among them. MI is adjusted for chance as follows~\citep{AMI}, which ranges between ${[0,1]}$ such that 1 indicates that two label assignments are equal:

\begin{equation}
	{AMI(C,C^{*})~=~} \dfrac{I() - \mathbf{E(I)}}{mean(H(C), H(C^{*})) - \mathbf{E(I)}}
\end{equation}

\textbf{Averaged Normalized Mutual Information:} As defined in~\citep{hypermeter}, in a given ensemble of clusterings ${K}$, we can obtain the single clustering ${C}$ information shared with the whole set of ensembles with the size of ${m}$ by:

\begin{equation}
	{ANMI(C,K)~=~} \dfrac{1}{m} \sum_{i~=~1}^{m}{NMI(C, C_i)} 
\end{equation}

\section{Distance Metrics}
\textbf{Jensen-Shannon Divergence (JSD):} Symmetric results and finite values are given when computing probability vectors defined as considering U and V are the vectors of different distributions. ${D}$ is the Kullback–Leibler divergence, and ${m}$ is the mean of the vectors computed pointwise. Kullback–Leibler measures the distance between two vectors by calculating entropy, and each vector value is summed to 1:

\begin{equation}
	JSD: \quad \sqrt{\dfrac {D(u\vert\vert m) + D (v||m)}{2}}
	\label{js}
\end{equation}

\begin{equation}
	{D_{KL}(U||M)~=~} -\sum_{x~\in X} U(x)\log \dfrac{M(x)}{U(x)}
	\label{kl}
	 \quad 
\end{equation}
\citep{menendez1997jensen}

\printcredits
\section*{Acknowledgement}
Authors would like to thank Demir\"{o}ren Teknoloji A.S. for the support and providing data. 

\bibliographystyle{cas-model2-names}
\bibliography{bib}

\begin{thebibliography}{62}
\expandafter\ifx\csname natexlab\endcsname\relax\def\natexlab#1{#1}\fi
\providecommand{\url}[1]{\texttt{#1}}
\providecommand{\href}[2]{#2}
\providecommand{\path}[1]{#1}
\providecommand{\DOIprefix}{doi:}
\providecommand{\ArXivprefix}{arXiv:}
\providecommand{\URLprefix}{URL: }
\providecommand{\Pubmedprefix}{pmid:}
\providecommand{\doi}[1]{\href{http://dx.doi.org/#1}{\path{#1}}}
\providecommand{\Pubmed}[1]{\href{pmid:#1}{\path{#1}}}
\providecommand{\bibinfo}[2]{#2}
\ifx\xfnm\relax \def\xfnm[#1]{\unskip,\space#1}\fi
\bibitem[{Akbari et~al.(2015)Akbari, Dahlan, Ibrahim and Alizadeh}]{ens_Akbari}
\bibinfo{author}{Akbari, E.}, \bibinfo{author}{Dahlan, H.M.},
  \bibinfo{author}{Ibrahim, R.}, \bibinfo{author}{Alizadeh, H.},
  \bibinfo{year}{2015}.
\newblock \bibinfo{title}{Hierarchical cluster ensemble selection}.
\newblock \bibinfo{journal}{Engineering Applications of Artificial
  Intelligence} \bibinfo{volume}{39}, \bibinfo{pages}{146--156}.
\bibitem[{Alizadeh et~al.(2014)Alizadeh, Minaei-Bidgoli and
  Parvin}]{ens_Alizadeh}
\bibinfo{author}{Alizadeh, H.}, \bibinfo{author}{Minaei-Bidgoli, B.},
  \bibinfo{author}{Parvin, H.}, \bibinfo{year}{2014}.
\newblock \bibinfo{title}{To improve the quality of cluster ensembles by
  selecting a subset of base clusters}.
\newblock \bibinfo{journal}{Journal of Experimental \& Theoretical Artificial
  Intelligence} \bibinfo{volume}{26}, \bibinfo{pages}{127--150}.
\bibitem[{Aras et~al.(2021)Aras, Makaro{\u{g}}lu, Demir and Cakir}]{NER}
\bibinfo{author}{Aras, G.}, \bibinfo{author}{Makaro{\u{g}}lu, D.},
  \bibinfo{author}{Demir, S.}, \bibinfo{author}{Cakir, A.},
  \bibinfo{year}{2021}.
\newblock \bibinfo{title}{An evaluation of recent neural sequence tagging
  models in turkish named entity recognition}.
\newblock \bibinfo{journal}{Expert Systems with Applications}
  \bibinfo{volume}{182}, \bibinfo{pages}{115049}.
\bibitem[{Arora et~al.(2016)Arora, Varshney et~al.}]{centroid-based}
\bibinfo{author}{Arora, P.}, \bibinfo{author}{Varshney, S.}, et~al.,
  \bibinfo{year}{2016}.
\newblock \bibinfo{title}{Analysis of k-means and k-medoids algorithm for big
  data}.
\newblock \bibinfo{journal}{Procedia Computer Science} \bibinfo{volume}{78},
  \bibinfo{pages}{507--512}.
\bibitem[{Azimi and Fern(2009)}]{ens_Azimi}
\bibinfo{author}{Azimi, J.}, \bibinfo{author}{Fern, X.}, \bibinfo{year}{2009}.
\newblock \bibinfo{title}{Adaptive cluster ensemble selection}, in:
  \bibinfo{booktitle}{Twenty-First International Joint Conference on Artificial
  Intelligence}.
\bibitem[{Banerjee and Dave(2004)}]{hopkins}
\bibinfo{author}{Banerjee, A.}, \bibinfo{author}{Dave, R.N.},
  \bibinfo{year}{2004}.
\newblock \bibinfo{title}{Validating clusters using the hopkins statistic}, in:
  \bibinfo{booktitle}{2004 IEEE International conference on fuzzy systems (IEEE
  Cat. No. 04CH37542)}, \bibinfo{organization}{IEEE}. pp.
  \bibinfo{pages}{149--153}.
\bibitem[{Bar-Gill et~al.(2021)Bar-Gill, Inbar and Reichman}]{Bar-Gill}
\bibinfo{author}{Bar-Gill, S.}, \bibinfo{author}{Inbar, Y.},
  \bibinfo{author}{Reichman, S.}, \bibinfo{year}{2021}.
\newblock \bibinfo{title}{The impact of social vs. nonsocial referring channels
  on online news consumption}.
\newblock \bibinfo{journal}{Management Science} \bibinfo{volume}{67},
  \bibinfo{pages}{2420--2447}.
\bibitem[{Benlian(2015)}]{Benlian}
\bibinfo{author}{Benlian, A.}, \bibinfo{year}{2015}.
\newblock \bibinfo{title}{Web personalization cues and their differential
  effects on user assessments of website value}.
\newblock \bibinfo{journal}{Journal of management information systems}
  \bibinfo{volume}{32}, \bibinfo{pages}{225--260}.
\bibitem[{Bogaard et~al.(2019a)Bogaard, Hollink, Wielemaker, Hardman and
  Van~Ossenbruggen}]{bogaard2019searching}
\bibinfo{author}{Bogaard, T.}, \bibinfo{author}{Hollink, L.},
  \bibinfo{author}{Wielemaker, J.}, \bibinfo{author}{Hardman, L.},
  \bibinfo{author}{Van~Ossenbruggen, J.}, \bibinfo{year}{2019}a.
\newblock \bibinfo{title}{Searching for old news: User interests and behavior
  within a national collection}, in: \bibinfo{booktitle}{Proceedings of the
  2019 Conference on Human Information Interaction and Retrieval}, pp.
  \bibinfo{pages}{113--121}.
\bibitem[{Bogaard et~al.(2019b)Bogaard, Hollink, Wielemaker, Hardman and
  Van~Ossenbruggen}]{TesselLaura}
\bibinfo{author}{Bogaard, T.}, \bibinfo{author}{Hollink, L.},
  \bibinfo{author}{Wielemaker, J.}, \bibinfo{author}{Hardman, L.},
  \bibinfo{author}{Van~Ossenbruggen, J.}, \bibinfo{year}{2019}b.
\newblock \bibinfo{title}{Searching for old news: User interests and behavior
  within a national collection}, in: \bibinfo{booktitle}{Proceedings of the
  2019 Conference on Human Information Interaction and Retrieval}, pp.
  \bibinfo{pages}{113--121}.
\bibitem[{Boongoen and Iam-On(2018)}]{ens_Boongoen}
\bibinfo{author}{Boongoen, T.}, \bibinfo{author}{Iam-On, N.},
  \bibinfo{year}{2018}.
\newblock \bibinfo{title}{Cluster ensembles: A survey of approaches with recent
  extensions and applications}.
\newblock \bibinfo{journal}{Computer Science Review} \bibinfo{volume}{28},
  \bibinfo{pages}{1--25}.
\bibitem[{Cali{\'n}ski and Harabasz(1974)}]{calinski}
\bibinfo{author}{Cali{\'n}ski, T.}, \bibinfo{author}{Harabasz, J.},
  \bibinfo{year}{1974}.
\newblock \bibinfo{title}{A dendrite method for cluster analysis}.
\newblock \bibinfo{journal}{Communications in Statistics-theory and Methods}
  \bibinfo{volume}{3}, \bibinfo{pages}{1--27}.
\bibitem[{Castellano et~al.(2013)Castellano, Fanelli and Torsello}]{Srivastava}
\bibinfo{author}{Castellano, G.}, \bibinfo{author}{Fanelli, A.M.},
  \bibinfo{author}{Torsello, M.A.}, \bibinfo{year}{2013}.
\newblock \bibinfo{title}{Web usage mining: discovering usage patterns for web
  applications}, in: \bibinfo{booktitle}{Advanced Techniques in Web
  Intelligence-2}. \bibinfo{publisher}{Springer}, pp. \bibinfo{pages}{75--104}.
\bibitem[{Catledge and Pitkow(1995)}]{Catledge}
\bibinfo{author}{Catledge, L.D.}, \bibinfo{author}{Pitkow, J.E.},
  \bibinfo{year}{1995}.
\newblock \bibinfo{title}{Characterizing browsing strategies in the world-wide
  web}.
\newblock \bibinfo{journal}{Computer Networks and ISDN systems}
  \bibinfo{volume}{27}, \bibinfo{pages}{1065--1073}.
\bibitem[{Chiou and Tucker(2017)}]{Tucker}
\bibinfo{author}{Chiou, L.}, \bibinfo{author}{Tucker, C.},
  \bibinfo{year}{2017}.
\newblock \bibinfo{title}{Content aggregation by platforms: The case of the
  news media}.
\newblock \bibinfo{journal}{Journal of Economics \& Management Strategy}
  \bibinfo{volume}{26}, \bibinfo{pages}{782--805}.
\bibitem[{Dellarocas et~al.(2013)Dellarocas, Katona and Rand}]{Dellarocas}
\bibinfo{author}{Dellarocas, C.}, \bibinfo{author}{Katona, Z.},
  \bibinfo{author}{Rand, W.}, \bibinfo{year}{2013}.
\newblock \bibinfo{title}{Media, aggregators, and the link economy: Strategic
  hyperlink formation in content networks}.
\newblock \bibinfo{journal}{Management science} \bibinfo{volume}{59},
  \bibinfo{pages}{2360--2379}.
\bibitem[{Dunn(1974)}]{Dunn}
\bibinfo{author}{Dunn, J.C.}, \bibinfo{year}{1974}.
\newblock \bibinfo{title}{Well-separated clusters and optimal fuzzy
  partitions}.
\newblock \bibinfo{journal}{Journal of cybernetics} \bibinfo{volume}{4},
  \bibinfo{pages}{95--104}.
\bibitem[{Fern and Brodley(2004)}]{ens_FernBrodley}
\bibinfo{author}{Fern, X.Z.}, \bibinfo{author}{Brodley, C.E.},
  \bibinfo{year}{2004}.
\newblock \bibinfo{title}{Solving cluster ensemble problems by bipartite graph
  partitioning}, in: \bibinfo{booktitle}{Proceedings of the twenty-first
  international conference on Machine learning}, p.~\bibinfo{pages}{36}.
\bibitem[{Fern and Lin(2008)}]{ens_Fern}
\bibinfo{author}{Fern, X.Z.}, \bibinfo{author}{Lin, W.}, \bibinfo{year}{2008}.
\newblock \bibinfo{title}{Cluster ensemble selection}.
\newblock \bibinfo{journal}{Statistical Analysis and Data Mining: The ASA Data
  Science Journal} \bibinfo{volume}{1}, \bibinfo{pages}{128--141}.
\bibitem[{Flaxman et~al.(2016)Flaxman, Goel and
  Rao}]{consiumingnews_Flaxman_2016}
\bibinfo{author}{Flaxman, S.}, \bibinfo{author}{Goel, S.},
  \bibinfo{author}{Rao, J.M.}, \bibinfo{year}{2016}.
\newblock \bibinfo{title}{Filter bubbles, echo chambers, and online news
  consumption}.
\newblock \bibinfo{journal}{Public opinion quarterly} \bibinfo{volume}{80},
  \bibinfo{pages}{298--320}.
\bibitem[{Freedman and Diaconis(1981)}]{FD_rule}
\bibinfo{author}{Freedman, D.}, \bibinfo{author}{Diaconis, P.},
  \bibinfo{year}{1981}.
\newblock \bibinfo{title}{On the histogram as a density estimator: L 2 theory}.
\newblock \bibinfo{journal}{Zeitschrift f{\"u}r Wahrscheinlichkeitstheorie und
  verwandte Gebiete} \bibinfo{volume}{57}, \bibinfo{pages}{453--476}.
\bibitem[{Grech and Clough(2016a)}]{DanielGrech}
\bibinfo{author}{Grech, D.}, \bibinfo{author}{Clough, P.},
  \bibinfo{year}{2016}a.
\newblock \bibinfo{title}{Investigating cluster stability when analyzing
  transaction logs}, in: \bibinfo{booktitle}{Proceedings of the 16th
  ACM/IEEE-CS on Joint Conference on Digital Libraries}, pp.
  \bibinfo{pages}{115--118}.
\bibitem[{Grech and Clough(2016b)}]{clustertend_DanielGrech}
\bibinfo{author}{Grech, D.}, \bibinfo{author}{Clough, P.},
  \bibinfo{year}{2016}b.
\newblock \bibinfo{title}{Investigating cluster stability when analyzing
  transaction logs}, in: \bibinfo{booktitle}{Proceedings of the 16th
  ACM/IEEE-CS on Joint Conference on Digital Libraries}, pp.
  \bibinfo{pages}{115--118}.
\bibitem[{Halkidi and Vazirgiannis(2001)}]{Sdbw_Halkidi}
\bibinfo{author}{Halkidi, M.}, \bibinfo{author}{Vazirgiannis, M.},
  \bibinfo{year}{2001}.
\newblock \bibinfo{title}{Clustering validity assessment: finding the optimal
  partitioning of a data set}, in: \bibinfo{booktitle}{Proceedings 2001 IEEE
  International Conference on Data Mining}, pp. \bibinfo{pages}{187--194}.
\newblock \DOIprefix\doi{10.1109/ICDM.2001.989517}.
\bibitem[{Handl et~al.(2005)Handl, Knowles and Kell}]{Handl}
\bibinfo{author}{Handl, J.}, \bibinfo{author}{Knowles, J.},
  \bibinfo{author}{Kell, D.B.}, \bibinfo{year}{2005}.
\newblock \bibinfo{title}{Computational cluster validation in post-genomic data
  analysis}.
\newblock \bibinfo{journal}{Bioinformatics} \bibinfo{volume}{21},
  \bibinfo{pages}{3201--3212}.
\bibitem[{Helfmann et~al.(2018)Helfmann, von Lindheim, Mollenhauer and
  Banisch}]{hypermeter}
\bibinfo{author}{Helfmann, L.}, \bibinfo{author}{von Lindheim, J.},
  \bibinfo{author}{Mollenhauer, M.}, \bibinfo{author}{Banisch, R.},
  \bibinfo{year}{2018}.
\newblock \bibinfo{title}{On hyperparameter search in cluster ensembles}.
\newblock \bibinfo{journal}{arXiv preprint arXiv:1803.11008} .
\bibitem[{Jain et~al.(1999)Jain, Murty and Flynn}]{cls_Jain}
\bibinfo{author}{Jain, A.K.}, \bibinfo{author}{Murty, M.N.},
  \bibinfo{author}{Flynn, P.J.}, \bibinfo{year}{1999}.
\newblock \bibinfo{title}{Data clustering: a review}.
\newblock \bibinfo{journal}{ACM computing surveys (CSUR)} \bibinfo{volume}{31},
  \bibinfo{pages}{264--323}.
\bibitem[{Jing et~al.(2015)Jing, Tian and Huang}]{Stratified}
\bibinfo{author}{Jing, L.}, \bibinfo{author}{Tian, K.}, \bibinfo{author}{Huang,
  J.Z.}, \bibinfo{year}{2015}.
\newblock \bibinfo{title}{Stratified feature sampling method for ensemble
  clustering of high dimensional data}.
\newblock \bibinfo{journal}{Pattern Recognition} \bibinfo{volume}{48},
  \bibinfo{pages}{3688--3702}.
\bibitem[{Kim and Lee(2003)}]{Sdbw_Kim}
\bibinfo{author}{Kim, Y.}, \bibinfo{author}{Lee, S.}, \bibinfo{year}{2003}.
\newblock \bibinfo{title}{A clustering validity assessment index}, in:
  \bibinfo{booktitle}{Pacific-Asia Conference on Knowledge Discovery and Data
  Mining}, \bibinfo{organization}{Springer}. pp. \bibinfo{pages}{602--608}.
\bibitem[{K{\"o}ster et~al.(2021)K{\"o}ster, Matt and Hess}]{Koster}
\bibinfo{author}{K{\"o}ster, A.}, \bibinfo{author}{Matt, C.},
  \bibinfo{author}{Hess, T.}, \bibinfo{year}{2021}.
\newblock \bibinfo{title}{Do all roads lead to rome? exploring the relationship
  between social referrals, referral propensity and stickiness to
  video-on-demand websites}.
\newblock \bibinfo{journal}{Business \& Information Systems Engineering}
  \bibinfo{volume}{63}, \bibinfo{pages}{349--366}.
\bibitem[{Kuncheva and Vetrov(2006)}]{ensembles_Kuncheva}
\bibinfo{author}{Kuncheva, L.I.}, \bibinfo{author}{Vetrov, D.P.},
  \bibinfo{year}{2006}.
\newblock \bibinfo{title}{Evaluation of stability of k-means cluster ensembles
  with respect to random initialization}.
\newblock \bibinfo{journal}{IEEE transactions on pattern analysis and machine
  intelligence} \bibinfo{volume}{28}, \bibinfo{pages}{1798--1808}.
\bibitem[{Li et~al.(2007)Li, Ding and Jordan}]{ensembles_DingJordan}
\bibinfo{author}{Li, T.}, \bibinfo{author}{Ding, C.}, \bibinfo{author}{Jordan,
  M.I.}, \bibinfo{year}{2007}.
\newblock \bibinfo{title}{Solving consensus and semi-supervised clustering
  problems using nonnegative matrix factorization}, in:
  \bibinfo{booktitle}{Seventh IEEE International Conference on Data Mining
  (ICDM 2007)}, \bibinfo{organization}{IEEE}. pp. \bibinfo{pages}{577--582}.
\bibitem[{Lin et~al.(2014)Lin, Xie, Guan, Li and Li}]{lin2014personalized}
\bibinfo{author}{Lin, C.}, \bibinfo{author}{Xie, R.}, \bibinfo{author}{Guan,
  X.}, \bibinfo{author}{Li, L.}, \bibinfo{author}{Li, T.},
  \bibinfo{year}{2014}.
\newblock \bibinfo{title}{Personalized news recommendation via implicit social
  experts}.
\newblock \bibinfo{journal}{Information Sciences} \bibinfo{volume}{254},
  \bibinfo{pages}{1--18}.
\bibitem[{Liu et~al.(2010a)Liu, Dolan and Pedersen}]{9}
\bibinfo{author}{Liu, J.}, \bibinfo{author}{Dolan, P.},
  \bibinfo{author}{Pedersen, E.R.}, \bibinfo{year}{2010}a.
\newblock \bibinfo{title}{Personalized news recommendation based on click
  behavior}, in: \bibinfo{booktitle}{Proceedings of the 15th international
  conference on Intelligent user interfaces}, pp. \bibinfo{pages}{31--40}.
\bibitem[{Liu et~al.(2010b)Liu, Li, Xiong, Gao and
  Wu}]{Clustering_Validation_Measures}
\bibinfo{author}{Liu, Y.}, \bibinfo{author}{Li, Z.}, \bibinfo{author}{Xiong,
  H.}, \bibinfo{author}{Gao, X.}, \bibinfo{author}{Wu, J.},
  \bibinfo{year}{2010}b.
\newblock \bibinfo{title}{Understanding of internal clustering validation
  measures}, in: \bibinfo{booktitle}{2010 IEEE international conference on data
  mining}, \bibinfo{organization}{IEEE}. pp. \bibinfo{pages}{911--916}.
\bibitem[{Makaro{\u{g}}lu et~al.(2019)Makaro{\u{g}}lu, {\c{C}}ak{\i}r and
  Kocaba{\c{s}}}]{Makaroglu}
\bibinfo{author}{Makaro{\u{g}}lu, D.}, \bibinfo{author}{{\c{C}}ak{\i}r, A.},
  \bibinfo{author}{Kocaba{\c{s}}, K.}, \bibinfo{year}{2019}.
\newblock \bibinfo{title}{Social media and clickstream analysis in turkish news
  with apache spark}, in: \bibinfo{booktitle}{International Conference on
  Intelligent and Fuzzy Systems}, \bibinfo{organization}{Springer}. pp.
  \bibinfo{pages}{221--228}.
\bibitem[{Mao and Zhang(2015)}]{Mao}
\bibinfo{author}{Mao, E.}, \bibinfo{author}{Zhang, J.}, \bibinfo{year}{2015}.
\newblock \bibinfo{title}{What drives consumers to click on social media ads?
  the roles of content, media, and individual factors}, in:
  \bibinfo{booktitle}{2015 48th Hawaii International Conference on System
  Sciences}, \bibinfo{organization}{IEEE}. pp. \bibinfo{pages}{3405--3413}.
\bibitem[{McConville et~al.(2021)McConville, Santos-Rodriguez, Piechocki and
  Craddock}]{N2D}
\bibinfo{author}{McConville, R.}, \bibinfo{author}{Santos-Rodriguez, R.},
  \bibinfo{author}{Piechocki, R.J.}, \bibinfo{author}{Craddock, I.},
  \bibinfo{year}{2021}.
\newblock \bibinfo{title}{N2d:(not too) deep clustering via clustering the
  local manifold of an autoencoded embedding}, in: \bibinfo{booktitle}{2020
  25th International Conference on Pattern Recognition (ICPR)},
  \bibinfo{organization}{IEEE}. pp. \bibinfo{pages}{5145--5152}.
\bibitem[{McInnes et~al.(2017)McInnes, Healy and Astels}]{hdbscan}
\bibinfo{author}{McInnes, L.}, \bibinfo{author}{Healy, J.},
  \bibinfo{author}{Astels, S.}, \bibinfo{year}{2017}.
\newblock \bibinfo{title}{hdbscan: Hierarchical density based clustering}.
\newblock \bibinfo{journal}{Journal of Open Source Software}
  \bibinfo{volume}{2}, \bibinfo{pages}{205}.
\bibitem[{McInnes et~al.(2020)McInnes, Healy and Melville}]{umap}
\bibinfo{author}{McInnes, L.}, \bibinfo{author}{Healy, J.},
  \bibinfo{author}{Melville, J.}, \bibinfo{year}{2020}.
\newblock \bibinfo{title}{Umap: uniform manifold approximation and projection
  for dimension reduction} .
\bibitem[{Men{\'e}ndez et~al.(1997)Men{\'e}ndez, Pardo, Pardo and
  Pardo}]{menendez1997jensen}
\bibinfo{author}{Men{\'e}ndez, M.}, \bibinfo{author}{Pardo, J.},
  \bibinfo{author}{Pardo, L.}, \bibinfo{author}{Pardo, M.},
  \bibinfo{year}{1997}.
\newblock \bibinfo{title}{The jensen-shannon divergence}.
\newblock \bibinfo{journal}{Journal of the Franklin Institute}
  \bibinfo{volume}{334}, \bibinfo{pages}{307--318}.
\bibitem[{M{\"o}ller et~al.(2020)M{\"o}ller, van~de Velde, Merten and
  Puschmann}]{Moller}
\bibinfo{author}{M{\"o}ller, J.}, \bibinfo{author}{van~de Velde, R.N.},
  \bibinfo{author}{Merten, L.}, \bibinfo{author}{Puschmann, C.},
  \bibinfo{year}{2020}.
\newblock \bibinfo{title}{Explaining online news engagement based on browsing
  behavior: Creatures of habit?}
\newblock \bibinfo{journal}{Social Science Computer Review}
  \bibinfo{volume}{38}, \bibinfo{pages}{616--632}.
\bibitem[{Pividori et~al.(2016)Pividori, Stegmayer and Milone}]{ens_Pividori}
\bibinfo{author}{Pividori, M.}, \bibinfo{author}{Stegmayer, G.},
  \bibinfo{author}{Milone, D.H.}, \bibinfo{year}{2016}.
\newblock \bibinfo{title}{Diversity control for improving the analysis of
  consensus clustering}.
\newblock \bibinfo{journal}{Information Sciences} \bibinfo{volume}{361},
  \bibinfo{pages}{120--134}.
\bibitem[{Rodr{\'\i}guez-Fern{\'a}ndez
  et~al.(2017)Rodr{\'\i}guez-Fern{\'a}ndez, Men{\'e}ndez and
  Camacho}]{Fernandez_clustering}
\bibinfo{author}{Rodr{\'\i}guez-Fern{\'a}ndez, V.},
  \bibinfo{author}{Men{\'e}ndez, H.D.}, \bibinfo{author}{Camacho, D.},
  \bibinfo{year}{2017}.
\newblock \bibinfo{title}{A study on performance metrics and clustering methods
  for analyzing behavior in uav operations}.
\newblock \bibinfo{journal}{Journal of Intelligent \& Fuzzy Systems}
  \bibinfo{volume}{32}, \bibinfo{pages}{1307--1319}.
\bibitem[{Rousseeuw(1987)}]{Silhouette}
\bibinfo{author}{Rousseeuw, P.J.}, \bibinfo{year}{1987}.
\newblock \bibinfo{title}{Silhouettes: a graphical aid to the interpretation
  and validation of cluster analysis}.
\newblock \bibinfo{journal}{Journal of computational and applied mathematics}
  \bibinfo{volume}{20}, \bibinfo{pages}{53--65}.
\bibitem[{Schubert et~al.(2017)Schubert, Sander, Ester, Kriegel and
  Xu}]{dbscan}
\bibinfo{author}{Schubert, E.}, \bibinfo{author}{Sander, J.},
  \bibinfo{author}{Ester, M.}, \bibinfo{author}{Kriegel, H.P.},
  \bibinfo{author}{Xu, X.}, \bibinfo{year}{2017}.
\newblock \bibinfo{title}{Dbscan revisited, revisited: why and how you should
  (still) use dbscan}.
\newblock \bibinfo{journal}{ACM Transactions on Database Systems (TODS)}
  \bibinfo{volume}{42}, \bibinfo{pages}{1--21}.
\bibitem[{Strehl and Ghosh(2002)}]{ensembles_Ghosh}
\bibinfo{author}{Strehl, A.}, \bibinfo{author}{Ghosh, J.},
  \bibinfo{year}{2002}.
\newblock \bibinfo{title}{Cluster ensembles---a knowledge reuse framework for
  combining multiple partitions}.
\newblock \bibinfo{journal}{Journal of machine learning research}
  \bibinfo{volume}{3}, \bibinfo{pages}{583--617}.
\bibitem[{Su and Chen(2015)}]{SuChen}
\bibinfo{author}{Su, Q.}, \bibinfo{author}{Chen, L.}, \bibinfo{year}{2015}.
\newblock \bibinfo{title}{A method for discovering clusters of e-commerce
  interest patterns using click-stream data}.
\newblock \bibinfo{journal}{electronic commerce research and applications}
  \bibinfo{volume}{14}, \bibinfo{pages}{1--13}.
\bibitem[{Thorson and Wells(2016)}]{ThorsonWells}
\bibinfo{author}{Thorson, K.}, \bibinfo{author}{Wells, C.},
  \bibinfo{year}{2016}.
\newblock \bibinfo{title}{Curated flows: A framework for mapping media exposure
  in the digital age}.
\newblock \bibinfo{journal}{Communication Theory} \bibinfo{volume}{26},
  \bibinfo{pages}{309--328}.
\bibitem[{Tong and Tan(2009)}]{Sdbw_Tong}
\bibinfo{author}{Tong, J.}, \bibinfo{author}{Tan, H.}, \bibinfo{year}{2009}.
\newblock \bibinfo{title}{Clustering validity based on the improved s\_dbw*
  index}.
\newblock \bibinfo{journal}{Journal of Electronics (China)}
  \bibinfo{volume}{26}, \bibinfo{pages}{258--264}.
\bibitem[{Varia et~al.(2014)Varia, Mathew et~al.}]{AWS}
\bibinfo{author}{Varia, J.}, \bibinfo{author}{Mathew, S.}, et~al.,
  \bibinfo{year}{2014}.
\newblock \bibinfo{title}{Overview of amazon web services}.
\newblock \bibinfo{journal}{Amazon Web Services} \bibinfo{volume}{105}.
\bibitem[{Vinh et~al.(2010)Vinh, Epps and Bailey}]{AMI}
\bibinfo{author}{Vinh, N.X.}, \bibinfo{author}{Epps, J.},
  \bibinfo{author}{Bailey, J.}, \bibinfo{year}{2010}.
\newblock \bibinfo{title}{Information theoretic measures for clusterings
  comparison: Variants, properties, normalization and correction for chance}.
\newblock \bibinfo{journal}{The Journal of Machine Learning Research}
  \bibinfo{volume}{11}, \bibinfo{pages}{2837--2854}.
\bibitem[{Wang et~al.(2013)Wang, Konolige, Wilson, Wang, Zheng and
  Zhao}]{WangKonolige}
\bibinfo{author}{Wang, G.}, \bibinfo{author}{Konolige, T.},
  \bibinfo{author}{Wilson, C.}, \bibinfo{author}{Wang, X.},
  \bibinfo{author}{Zheng, H.}, \bibinfo{author}{Zhao, B.Y.},
  \bibinfo{year}{2013}.
\newblock \bibinfo{title}{You are how you click: Clickstream analysis for sybil
  detection}, in: \bibinfo{booktitle}{22nd $\{$USENIX$\}$ Security Symposium
  ($\{$USENIX$\}$ Security 13)}, pp. \bibinfo{pages}{241--256}.
\bibitem[{Wang et~al.(2016)Wang, Zhang, Tang, Zheng and Zhao}]{WangZhang}
\bibinfo{author}{Wang, G.}, \bibinfo{author}{Zhang, X.}, \bibinfo{author}{Tang,
  S.}, \bibinfo{author}{Zheng, H.}, \bibinfo{author}{Zhao, B.Y.},
  \bibinfo{year}{2016}.
\newblock \bibinfo{title}{Unsupervised clickstream clustering for user behavior
  analysis}, in: \bibinfo{booktitle}{Proceedings of the 2016 CHI conference on
  human factors in computing systems}, pp. \bibinfo{pages}{225--236}.
\bibitem[{Wegmann et~al.(2021)Wegmann, Zipperling, Hillenbrand and
  Fleischer}]{evaluation_review}
\bibinfo{author}{Wegmann, M.}, \bibinfo{author}{Zipperling, D.},
  \bibinfo{author}{Hillenbrand, J.}, \bibinfo{author}{Fleischer, J.},
  \bibinfo{year}{2021}.
\newblock \bibinfo{title}{A review of systematic selection of clustering
  algorithms and their evaluation}.
\newblock \bibinfo{journal}{arXiv preprint arXiv:2106.12792} .
\bibitem[{Wells and Thorson(2017)}]{WellsChris}
\bibinfo{author}{Wells, C.}, \bibinfo{author}{Thorson, K.},
  \bibinfo{year}{2017}.
\newblock \bibinfo{title}{Combining big data and survey techniques to model
  effects of political content flows in facebook}.
\newblock \bibinfo{journal}{Social Science Computer Review}
  \bibinfo{volume}{35}, \bibinfo{pages}{33--52}.
\bibitem[{Xu et~al.(2014)Xu, Forman, Kim and Van~Ittersum}]{XuForman}
\bibinfo{author}{Xu, J.}, \bibinfo{author}{Forman, C.}, \bibinfo{author}{Kim,
  J.B.}, \bibinfo{author}{Van~Ittersum, K.}, \bibinfo{year}{2014}.
\newblock \bibinfo{title}{News media channels: Complements or substitutes?
  evidence from mobile phone usage}.
\newblock \bibinfo{journal}{Journal of Marketing} \bibinfo{volume}{78},
  \bibinfo{pages}{97--112}.
\bibitem[{Yanatma(2018)}]{TurkeyReport}
\bibinfo{author}{Yanatma, S.}, \bibinfo{year}{2018}.
\newblock \bibinfo{title}{Reuters institute digital news report 2018--turkey
  supplementary report} .
\bibitem[{Yang et~al.(2017)Yang, Li, Zhou and Xiao}]{ens_Yang}
\bibinfo{author}{Yang, F.}, \bibinfo{author}{Li, T.}, \bibinfo{author}{Zhou,
  Q.}, \bibinfo{author}{Xiao, H.}, \bibinfo{year}{2017}.
\newblock \bibinfo{title}{Cluster ensemble selection with constraints}.
\newblock \bibinfo{journal}{Neurocomputing} \bibinfo{volume}{235},
  \bibinfo{pages}{59--70}.
\bibitem[{Yeo and Johnson(2000)}]{yeo_johnson}
\bibinfo{author}{Yeo, I.K.}, \bibinfo{author}{Johnson, R.A.},
  \bibinfo{year}{2000}.
\newblock \bibinfo{title}{A new family of power transformations to improve
  normality or symmetry}.
\newblock \bibinfo{journal}{Biometrika} \bibinfo{volume}{87},
  \bibinfo{pages}{954--959}.
\bibitem[{Zhang and Kamps(2010)}]{ZhangKamps}
\bibinfo{author}{Zhang, J.}, \bibinfo{author}{Kamps, J.}, \bibinfo{year}{2010}.
\newblock \bibinfo{title}{Search log analysis of user stereotypes, information
  seeking behavior, and contextual evaluation}, in:
  \bibinfo{booktitle}{Proceedings of the third symposium on Information
  interaction in context}, pp. \bibinfo{pages}{245--254}.
\bibitem[{Zheng et~al.(2013)Zheng, Li, Hong and Li}]{zheng2013penetrate}
\bibinfo{author}{Zheng, L.}, \bibinfo{author}{Li, L.}, \bibinfo{author}{Hong,
  W.}, \bibinfo{author}{Li, T.}, \bibinfo{year}{2013}.
\newblock \bibinfo{title}{Penetrate: Personalized news recommendation using
  ensemble hierarchical clustering}.
\newblock \bibinfo{journal}{Expert Systems with Applications}
  \bibinfo{volume}{40}, \bibinfo{pages}{2127--2136}.

\end{thebibliography}

\end{document}